\documentclass[aps,prl,reprint,showpacs]{revtex4-1}
\usepackage{graphicx}
\usepackage{amsmath}
\usepackage{amssymb}
\usepackage{dsfont}
\usepackage{hyperref}
\usepackage{fancyhdr}
\newcommand{\bra}[1]{\left< #1 \right|}
\newcommand{\ket}[1]{\left|#1\right>}

\renewcommand{\vec}[1]{\mathbf{#1}}
\newcounter{Bild}

\begin{document}
\title{Hybrid spin and valley quantum computing with singlet-triplet qubits}
\author{Niklas Rohling, Maximilian Russ, and Guido Burkard}
\affiliation{Department of Physics, University of Konstanz, D-78457 Konstanz, Germany}


\begin{abstract}
The valley degree of freedom in the electronic band structure of
silicon, graphene, and other materials is often
considered to be an obstacle for quantum computing (QC) based on electron
spins in quantum dots.
Here we show that control over the valley state opens new possibilities
for quantum information processing.
Combining qubits encoded in the singlet-triplet subspace
of spin and valley states allows for universal QC
using a universal two-qubit gate directly provided by the exchange interaction.
We show how spin and valley qubits can be separated in order to allow
for single-qubit rotations.
\end{abstract}

\maketitle
\thispagestyle{fancy}

\setcounter{Bild}{1}

\textit{Introduction.}---An alternative to single-spin qubits in quantum dots \cite{LoDi1998}
is to encode each qubit in a
double quantum dot (DQD) within the two-dimensional subspace spanned by the spin
singlet $\ket{S}=(\ket{01}-\ket{10})/\sqrt{2}$
and the triplet
$\ket{T_0}=(\ket{01}+\ket{10})/\sqrt{2}$
\cite{kloeffel_loss_review_2013},
with $\ket{0}=\ket{\uparrow}$, $\ket{1}=\ket{\downarrow}$.
For this singlet-triplet qubit, single-qubit rotations can be realized by the exchange
interaction \cite{petta,foletti2009} and a gradient in the magnetic field \cite{foletti2009};
two-qubit gates have been proposed based on exchange interaction
\cite{Levy2002,Benjamin_pra2001,Klinovaja_prb2012,PhysRevB.86.205306}
or electrostatic coupling \cite{Taylor_natphys,Stepanenko2007,PhysRevB.84.155329}.
An electrostatically controlled entangling gate has
been realized experimentally \cite{PhysRevLett.107.030506,Shulman}.
Quantum computing (QC) with single-spin and/or single-valley qubits requires rotations of
single-valley and/or single-spin degrees of freedom (DOFs) \cite{rohling_burkard}.
In this Letter, we show that, by extending the concept of $S$-$T_0$ qubits to electrons with
spin and valley DOFs, the exchange interaction together
with Zeeman gradients directly provides universal QC.
Single-qubit rotations are feasible when the spin and the valley
$S$-$T_0$ qubits are stored in separated DQDs,
whereas in a dual-used DQD, i.e., containing the $S$-$T_0$ spin and valley qubits,
a two-qubit gate is obtained by exchange interaction \cite{rohling_burkard}.
Here, we focus on operations separating and bringing together spin and valley qubits,
allowing for a universal set of single- and two-qubit operations in a quantum register.

The valley DOF describes the existence of nonequivalent minima (maxima)
in the conduction (valence) band in several materials such as
silicon \cite{zwanenburg}, graphene \cite{castro_neto}, carbon nanotubes
\cite{kuemmeth}, aluminum arsenide \cite{shayegan}, or
transition metal dichalcogenide monolayers \cite{xiao,kormanyos}.
A twofold valley degeneracy can be considered as a qubit.
In particular, silicon-based heterostructures have aroused much interest as hosts
for electron spin qubits \cite{zwanenburg}
due to long relaxation \cite{feher_gere} and coherence
\cite{tyryshkin} times.
Valley states in silicon structures have
been under intense theoretical
\cite{boykin:115,nestoklon,chutia,PhysRevB.80.081305,PhysRevB.82.155312,PhysRevB.82.245314,PhysRevB.84.155320,baena,drumm,de_michelis,gamble2012,gamble2013,tahan_joynt_2013,jiang,genetic_design,dusko}
and experimental
\cite{xiao:032103,borselli:123118,borselli:063109,lim,simmons,PhysRevB.86.115319,yang2013,PhysRevLett.111.046801,kott,roche,yuan,yamahata,hao,goswami}
investigation recently. 
The valley degeneracy is often considered problematic for spin QC
\cite{eriksson,trauzettel}, and valley splitting is used
to achieve pure spin exchange interaction \cite{maune}. 
Also, theories for the manipulation of valley qubits have been developed
for carbon nanotubes \cite{PhysRevLett.106.086801},
graphene \cite{PhysRevB.84.195463,PhysRevB.88.125422},
and silicon \cite{PhysRevB.82.205315,PhysRevLett.108.126804,PhysRevB.86.035321}.
While these approaches consider valley and spin qubits separately, we
investigate a hybrid quantum register containing both spin and valley qubits.

\begin{figure}[t]
\includegraphics[width=.85\columnwidth]{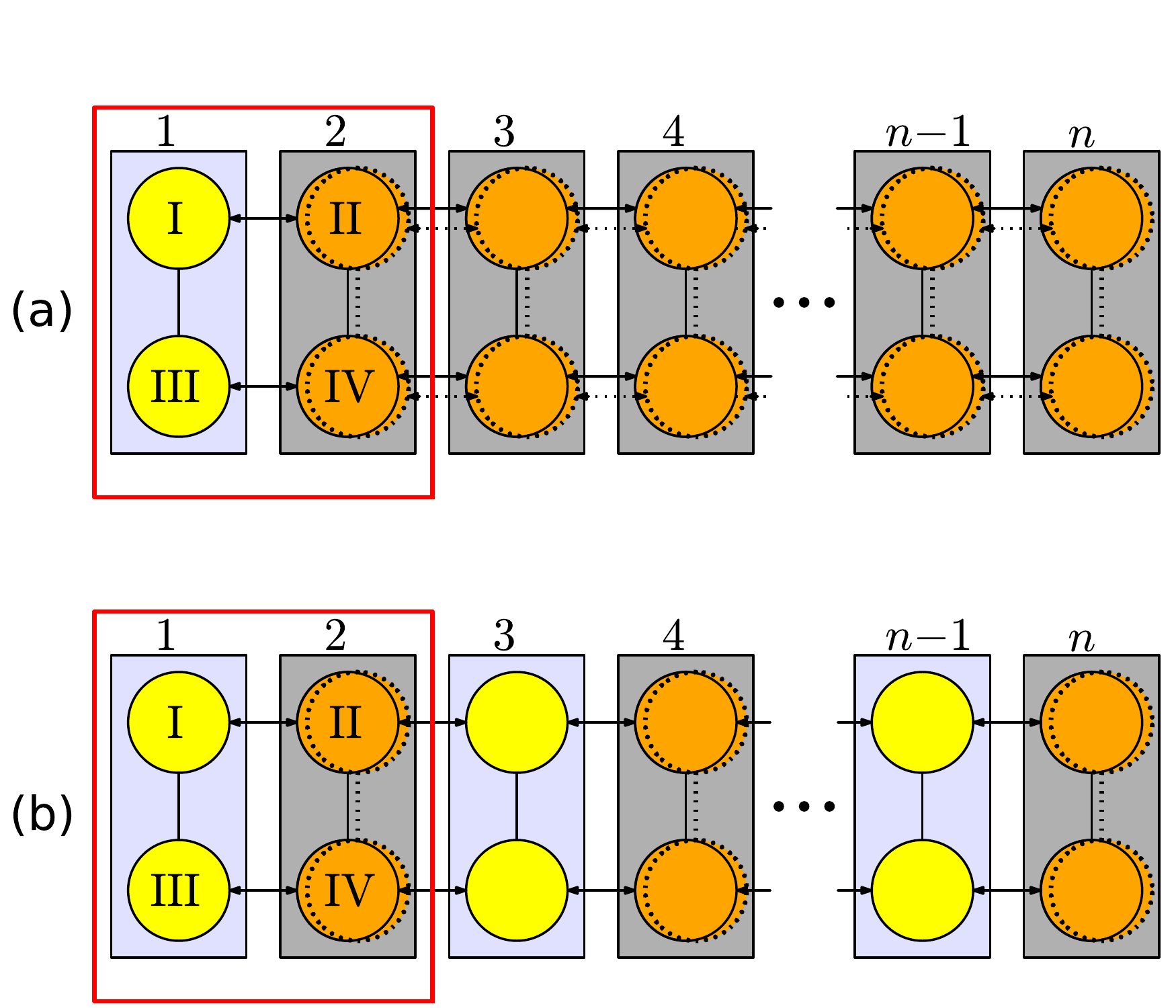} 
\caption{Two quantum registers for QC with spin and valley singlet-triplet qubits.
The single (yellow) circles denote dots with spin DOF only (valley degeneracy lifted)
hosting auxiliary spins,
whereas the double (orange) circles denote
dots with both spin and valley DOFs storing one spin and one valley $S$-$T_0$ qubit.
The lines within the DQDs represent the exchange interaction
(a single line for spin-only dots, solid and dotted lines for spin and valley),
which can realize single-qubit gates when the spin qubit is transferred to
the spin-only DQD and a two-qubit gate
when spin and valley qubits are in the same DQD.
The arrows between dots from neighboring DQDs represent ${\rm SWAP}$
operations,
interchanging the spin (solid arrow) or valley (dotted arrow) state between these dots.
Register (b) has more auxiliary dots; quantum gate sequences are shorter than in (a).}
\label{fig:register}
\end{figure}
\stepcounter{Bild}

Two different types of hybrid spin-valley singlet-triplet quantum
registers will be studied (Fig. \ref{fig:register}).
Both setups comprise two kinds of DQDs:
one (e.g.,dots I and III) with a spin DOF only
(simple yellow circles),
where the valley degeneracy does not exist or has been lifted
and another kind of DQD (e.g., II and IV) with both spin and twofold valley DOF
(double orange circles).
The elementary building block of the proposed
quantum register consists of two DQDs, one of each
kind (red rectangle in Fig.~\ref{fig:register}). 
We use the singlet $\ket{S}$ and  triplet $\ket{T_0}$ states of
spin and valley in DQD 2 as the logical qubits.
The spins in DQD 1 are spin polarized,
$\ket{\uparrow\uparrow}$ or $\ket{\downarrow\downarrow}$, 
which is needed for the single-qubit gates.
The exchange interaction between dots II and IV,
described in general by a Kugel-Khomskii Hamiltonian \cite{kugel_khomskii_1973,rohling_burkard},
already leads to a universal two-qubit gate between  the spin and the valley singlet-triplet
qubits in the same DQD \cite{rohling_burkard}.
Single-qubit gates for the logical qubits can be achieved by applying two 
spin-only SWAP gates, between I and II and between III and IV,
interchanging the spin $S$-$T_0$ qubit in DQD 2 with the polarized
ancilla spins in DQD 1, without affecting the valley state.
The single-qubit gates of the spin and the valley qubits can be
realized by exchange interaction and a spin or valley Zeeman gradient \cite{foletti2009,PhysRevLett.108.126804}. 
State preparation and measurements can be done for spatially separated spin and valley qubits.
Changing the detuning within the DQD maps exactly one state of the qubit to a state with both
electrons in one dot due to the Pauli exclusion principle \cite{kane_nature,vandersypen}.
The charge state can be detected by a quantum point contact \cite{petta}.
State preparation works in the opposite direction, starting from a ground state with both
electrons in one dot at large detuning.
Before explaining how to perform arbitrary quantum gates in the
registers, we propose a realization of an elementary gate
in our scheme, the spin-only SWAP gate.

\textit{Model.}---We consider the system consisting of dots I and II in Fig.~\ref{fig:register},
i.e., one dot with a spin DOF only and the other dot with spin and
twofold valley DOF.
We model the physics in this DQD by the Hamiltonian $H=H_0+H_T+H_V$.
The influence of the detuning $\varepsilon$ and the Coulomb repulsion energy between two electrons on the same site, $U$,
is given by $H_0 = \varepsilon(\hat n_1 - \hat n_2)/2 + U\sum_{i=1,2}\hat n_i(\hat n_i -1)/2$
with the number operators $\hat n_1=\sum_s\hat c_{1s}^\dagger\hat c_{1s}$
and $\hat n_2=\sum_{s,v}\hat c_{2sv}^\dagger \hat c_{2sv}$
where $\hat c_{1s}^{(\dagger)}$ annihilates (creates)
an electron with spin $s=\uparrow,\downarrow$
in dot I and $\hat c_{2sv}^{(\dagger)}$ annihilates (creates) an
electron with 
spin $s=\uparrow,\downarrow$ and valley $v=\pm$ in dot II.
Electron hopping is described by
$H_T = \sum_{s,v}t_v\hat c_{1s}^\dagger\hat c_{2sv} + h.c.$ and
the valley splitting by
$H_V = h\sum_s(\hat c_{2s+}^\dagger \hat c_{2s+}-\hat c_{2s-}^\dagger \hat c_{2s-}) \equiv h\sigma_z$,
where $\sigma_z$ is a Pauli matrix acting on the valley space.
The sums run over $s=\uparrow,\downarrow$
and $v=\pm$.
The valley-dependent hopping elements $t_\pm$ can be expressed by \cite{PhysRevB.82.205315}
$t_\pm = t(1\pm e^{i\varphi})/2$ with real parameters $t$ and $\varphi$,
and $h$ can be tuned in silicon by the electrically controlled confinement potential \cite{yang2013}
and in graphene by an out-of-plane magnetic field \cite{PhysRevB.79.085407}.
For two electrons in dots I and II, there are 15 possible states, one with (2,0),
six with (0,2), and eight with (1,1) charge distribution between dots I and II.
In the limit $t\ll|h+(U\pm\varepsilon)|,|h-(U\pm\varepsilon)|$,
a Schrieffer-Wolff transformation \cite{burkard_imamoglu} yields the
effective Hamiltonian \cite{Supplemental}
\begin{eqnarray}
\label{eqn:Heff}
 H_{\rm eff}& = & \left[(A\cos\varphi{+}B)\mathds{1} + (A{+}B\cos\varphi)\sigma_z + B\sin\varphi\,\sigma_y\right]P_S\nonumber\\
             &  &+ \tilde h\sigma_z
                +C\left[(1-\cos\varphi)\mathds{1}-\sin\varphi\,\sigma_y\right].               
\end{eqnarray}
Here, $P_S$ is the projector on the spin singlet
$\ket{S}=(\ket{\uparrow\downarrow}-\ket{\downarrow\uparrow})/\sqrt{2}$,
$A=4t^2Uh\varepsilon/[(h-U-\varepsilon)(h+U-\varepsilon)(h-U+\varepsilon)(h+U+\varepsilon)]$,
$B=A(h^2{+}\varepsilon^2{-}U^2)/2h\varepsilon$,
$\tilde h = h\{1+t^2(1{-}\cos\varphi)/[2(h{-}U{+}\varepsilon)(h{+}U{-}\varepsilon)]\}$, and
$C = t^2(U{-}\varepsilon)/[2(h{-}U{+}\varepsilon)(h{+}U{-}\varepsilon)]$.
Note that $H$ and $H_{\rm eff}$ are block diagonal in the spin
singlet-triplet basis, similar to the situation in Ref.~\cite{PhysRevB.82.205315}.
We aim at using the term proportional to $P_S$
to perform a spin-only ${\rm SWAP}$ gate, interchanging 
the spin information of dots I and II independently of the valley state.
We will show that this is possible despite the valley dependence of $H_{\rm eff}$.

\textit{Spin-only SWAP gate.}---For the valley-degenerate case $h=0$, 
Eq.~(\ref{eqn:Heff}) simplifies to the exchange Hamiltonian
\begin{equation}
 H^0_{\rm eff} = - JP_S^{}P_{k} -\tilde JP_{k^\perp},
\end{equation}
where
$J = 4t^2U/(U^2-\varepsilon^2)$ and $\tilde J = t^2/(U-\varepsilon)$.
The first term is proportional to $P_S$ 
and to the projector $P_k$ on the valley state
$\ket{k}=[(1+e^{-i\varphi})\ket{+}+(1-e^{-i\varphi})\ket{-}]/2$
occupied by an electron after hopping from dot I to dot II.
The second term is spin independent and $\propto P_{ k^\perp}=\ket{k^\perp}\bra{k^\perp}$,
where $\ket{k^\perp}=[(1-e^{-i\varphi})\ket{+}+(1+e^{-i\varphi})\ket{-}]/2$
is orthogonal to $\ket{k}$.
Both contributions originate from the Pauli principle:
virtual hopping between dots I and II is possible only if the participating
(0,2) or (2,0) state is antisymmetric.
Virtual hopping from (1,1) to (0,2) and back is $\propto t^2/(U-\varepsilon)$
and dominates for $\varepsilon\approx U$;
if the valley state in the right dot is $\ket{k^\perp}$, this channel is open, independently
of the spin states, as the valley state $\ket{k}$ used within the virtual hopping is empty.
If the valley state is $\ket{k}$, the spins have to be in a singlet to allow for
an antisymmetric (0,2) state.
Virtual hopping from (1,1) to (2,0) and back requires a valley state $\ket{k}$ in the right dot,
as $\ket{k^\perp}$ has no overlap with the state in the left dot; it
further requires the spins to be a singlet to obey Fermi-Dirac statistics.

For $|U+\varepsilon|\ll U$, $\tilde J$ can be neglected and the time evolution
$U_0(\phi)=\exp[-i\int_0^\tau d\tau' H^0_{\rm eff}(\tau')/\hbar]$ with $\phi=\int_0^\tau d\tau'J(\tau')/\hbar$
can be computed easily: $U_0(\phi)=\mathds{1}+(e^{i\phi}-1)P_SP_k$.
If the conditions $\tilde J\approx0$, $\varphi=\pi/2$, and controllability of
the valley splitting $h$ are fulfilled, we obtain a spin-only SWAP gate with the sequence
${\rm SWAP}\equiv\mathds{1}-2P_S=\sigma_zU_0(\pi)\sigma_zU_0(\pi)$.
Here, $h$ is turned off or at least made negligibly small during the exchange interaction but dominates
over exchange in between to realize the valley gate $\sigma_z$.

\begin{figure}
\includegraphics[width=1\columnwidth]{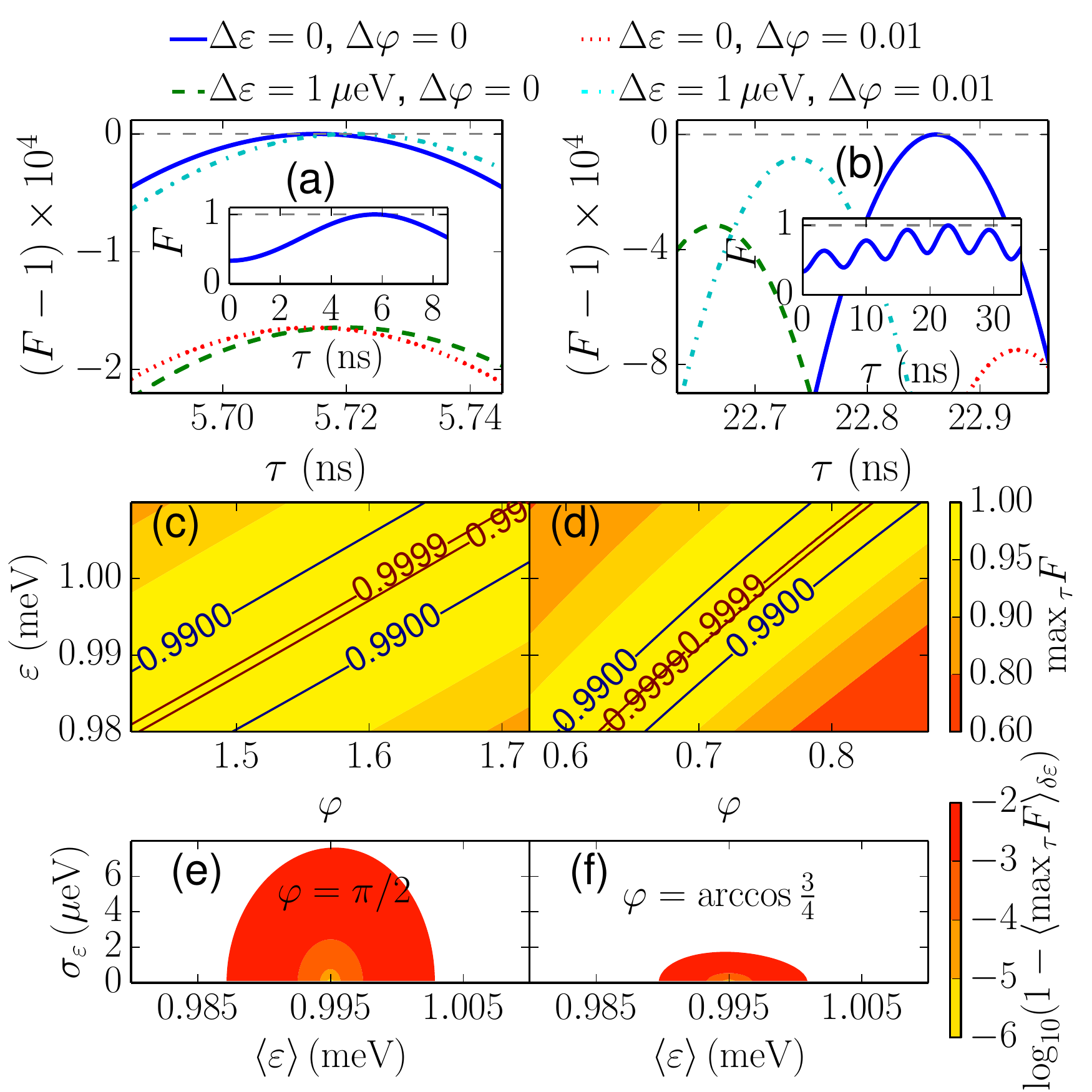} 
\caption{Fidelity $F$ of the time evolution operator $U_{\rm eff}(\tau)$
describing the exchange interaction between a spin-only dot and a
spin-valley dot with respect to thes spin SWAP gate according to Eq.~(\ref{eqn:Fidelity}).
The parameters are chosen to be $U=1$~meV, $h=0.1$~meV, and $t=6~\mu$eV.
(a),b) $F$ maximized over the gate time $\tau$ (scale bar).  
The detuning is chosen to be
$\varepsilon=\sqrt{U^2-h^2}+\Delta\varepsilon\approx0.995~{\rm meV}+\Delta\varepsilon$
and the phase in the hopping matrix elements,
$\varphi=\frac{\pi}{2}+\Delta\varphi$ (a) and
$\varphi=\arccos\frac{3}{4}+\Delta\varphi$ (b).
(c),d) Contour plots of maximal fidelity in dependence of $\varepsilon$ and $\varphi$.
The maximum over $\tau$ is taken numerically
by searching around the first (c) and fourth (d) local maximum.
At $\Delta\varphi=0$, $\Delta\varepsilon=0$, $F=1$ can be realized.
A shift in valley hopping phase, $\Delta\varphi$, can be compensated by adjusting $\varepsilon$.
(e),f) Averaged fidelity for a Gaussian distribution of $\varepsilon$ with
variance 
$\sigma_\varepsilon^2=\langle\varepsilon^2\rangle-\langle\varepsilon\rangle^2$.
The maximization over time is done for the value $\varepsilon=\langle\varepsilon\rangle$ at
the first (e) and fourth (f) local maximum.}
\label{fig:fidelity}
\stepcounter{Bild}
\end{figure}
\begin{figure}
\includegraphics[width=1\columnwidth]{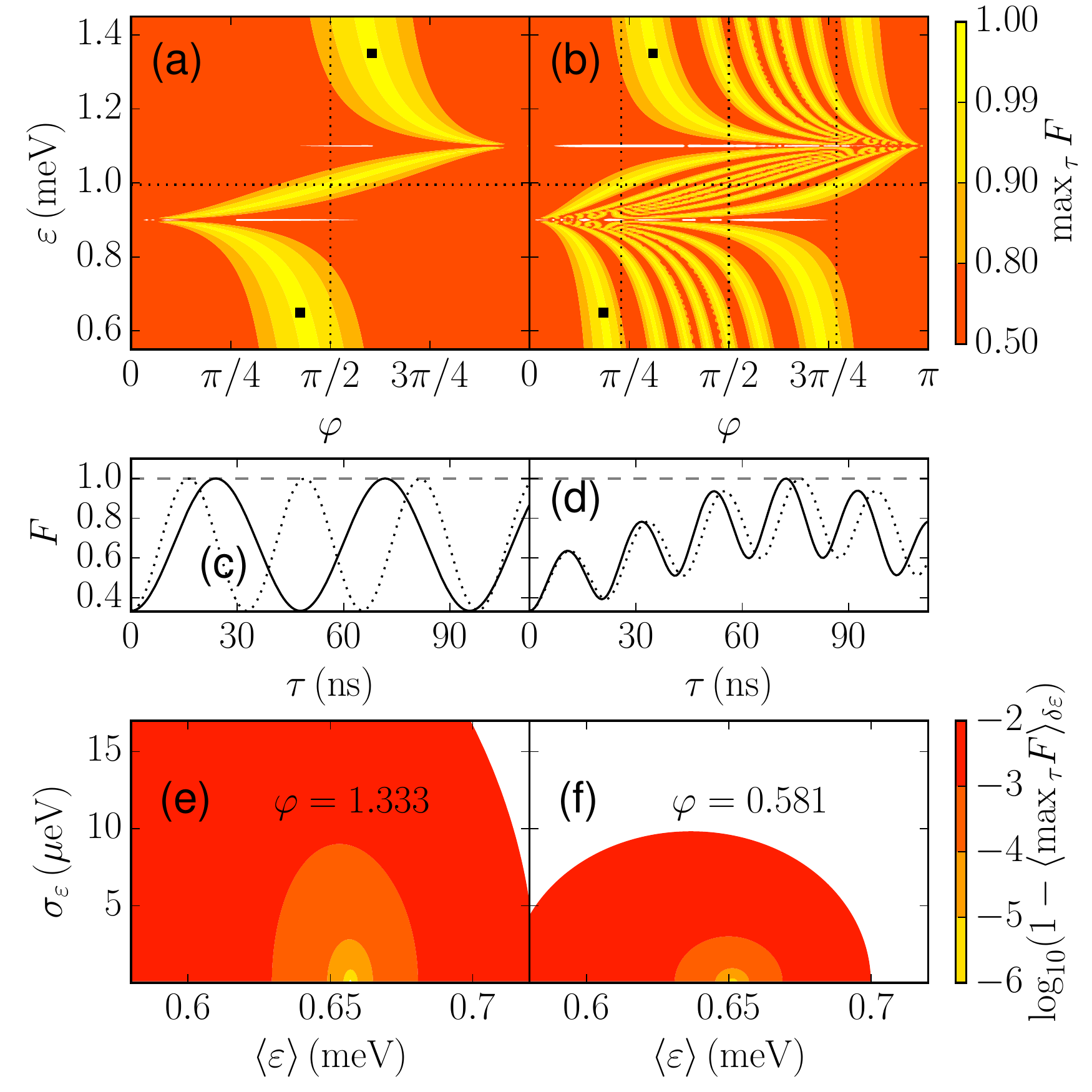} 
\caption{Fidelity $F$ for a broad range of
the detuning $\varepsilon$ and the phase $\varphi$  in the hopping
matrix elements, using the same parameters $U$, $h$, and $t$ as in Fig.~\ref{fig:fidelity}.
(a),b) $F$ maximized over the gate time $\tau$. 
We consider the first (a) and fourth (b) local maximum of $F$ as a function of $\tau$.
The black squares indicate the positions in parameter space for the time-dependent plots
(c,d), namely, $\varepsilon=1.35\,{\rm meV}$ for the solid lines and $\varepsilon=0.65\,{\rm meV}$
for the dotted lines in (c),d),
$\varphi=1.9$ (solid line), $\varphi=1.333$ (dotted line) in (c),
and $\varphi=0.972$ (solid line), $\varphi=0.581$ (dotted line) in (d).
The horizontal dotted lines in (a),b) represent $\varepsilon=\sqrt{U^2{-}h^2}$,
and the vertical dotted lines belong to values of $\varphi$ being $\frac{\pi}{2}$ (a),b) and
$\arccos\frac{3}{4}$, $\pi-\arccos\frac{3}{4}$ (b).
Note that the Schrieffer-Wolff transformation breaks down close to $\varepsilon=U\pm h$.
(e),f) Averaged fidelity for a Gaussian distribution of $\varepsilon$ with
$\sigma_\varepsilon^2=\langle\varepsilon^2\rangle-\langle\varepsilon\rangle^2$.
The maximum in time is determined for $\varepsilon=\langle\varepsilon\rangle$ at
the first (e) and fourth (f) local maximum.
The fidelity, $F$ is more robust at the first local maximum.}
\label{fig:fidelity2}
\end{figure}

If $\tilde h\sigma_z$ is the dominant contribution in
$H_{\rm eff}$, we find parameters which allow for a spin SWAP gate
(Figs.~\ref{fig:fidelity} and \ref{fig:fidelity2}).
We denote the time evolution according to a time-independent $H_{\rm eff}$
at time $\tau$ with $U_{\rm eff}(\tau)$ and consider the average gate fidelity \cite{Pedersen200747}
\begin{equation}
 F=\max_\alpha \frac{8+|\operatorname{Tr}[e^{i\alpha\sigma_z}U_{\rm eff}(\tau){\rm SWAP}^\dagger]|^2}{72}
\label{eqn:Fidelity}
\end{equation}
where we maximize over a $z$ rotation in valley space, which is
unimportant for the logical valley qubit in the singlet-triplet subspace.
For the spin SWAP gate applied between dots I and II and between
dots III and IV, the valley $z$ rotation can be different.
This difference is equivalent to a rotation of the $S$-$T_0$ valley qubit,
which can be corrected afterwards.
In the case $|A|,|B|,|C|\ll|\tilde h|$ considered here,
we determine the phase $\alpha$ for the maximum in
Eq.~(\ref{eqn:Fidelity}) analytically \cite{Supplemental}.
Figure \ref{fig:fidelity} shows $F$ for a detuning around $\varepsilon=\sqrt{U^2{-}h^2}$,
where the parameter $B=0$.
If furthermore $C$ is negligible, we obtain $U_{\rm eff}(\tau_n)={\rm SWAP}$
for $\varphi=\arccos\frac{n}{n+1}$ and $\tau_n=(n+1)\pi\hbar/A$.
In Fig.~\ref{fig:fidelity}, the situation is shown for $n=0$ and $n=3$.
Figure \ref{fig:fidelity2} reveals that, also for other parameter regimes,
$U_{\rm eff}(\tau)$ can be the spin SWAP gate with a high fidelity.
The time $\tau$ where the maximal fidelity is reached is determined numerically.
We include quasistatic charge noise, which was found to be important
for GaAs DQDs \cite{PhysRevLett110.146804}, by averaging $F$
over a Gaussian distribution of $\varepsilon$.
We find that for the fluctuating bias $\varepsilon$ with standard deviation
$\sigma_\varepsilon = \sqrt{\langle\varepsilon^2\rangle-\langle\varepsilon\rangle^2}\approx\mu$eV
the fidelity can be as high as $F\gtrsim0.9999$
[see Figs. \ref{fig:fidelity}(e), \ref{fig:fidelity}(f), Fig. \ref{fig:fidelity2}(e), and \ref{fig:fidelity2}(f)].
We expect the noise sensitivity of $U_{\rm eff}(\tau)$ to be similar
to exchange gates with spin DOF only \cite{Supplemental}.
Spin singlet-triplet qubits in silicon also suffer from coupling to nuclear spins,
but recent work concludes that charge noise is the dominating source
of dephasing \cite{APL102_232108}.

\textit{Hybrid quantum register.}---Now we consider the entire
register of $n$ DQDs built in two different ways
(Fig.~\ref{fig:register}) and prove that universal QC is possible in these 
two registers.
It is sufficient to show that single-qubit 
gates for every qubit and a universal two-qubit gate between arbitrary qubits can be performed \cite{divincenzo1995}.

In the register Fig.~\ref{fig:register} (a), there is only one spin-only DQD, at position 1,
and $(n-1)$ DQDs with spin and valley DOFs.
Single-qubit gates for the qubits in the $k$th DQD, $1< k\le n$ can be performed by first transferring
these qubits, both spin and valley, to DQD 2.
This is done by applying ${\rm SWAP}_{\rm spin}\otimes{\rm SWAP}_{\rm valley}$ gates
in the upper and the lower row of the register which allows interchanging the information of
DQD $k$ with $k-1$ and so on.
The ${\rm SWAP}_{\rm spin}\otimes{\rm SWAP}_{\rm valley}$ gate is provided directly
from the exchange interaction \cite{rohling_burkard}.
Second, the spin-only ${\rm SWAP}$ gate transfers the spin qubit into DQD 1, and
in return the polarized ancilla spins into DQD 2.
Now single-qubit operations can be performed for the spin $S$-$T_0$ qubit in DQD 1
and for the valley in DQD 2.
The universal two-qubit gate between qubits in DQD $k$ and in DQD $m$ require
transferring the qubits from these dots to DQDs 2 and 3.
Then applying the spin-only ${\rm SWAP}$ moves the spin qubit from DQD 2 to DQD 1.
The ${\rm SWAP}_{\rm spin}\otimes{\rm SWAP}_{\rm valley}$ gate between DQDs 2 and 3 and
another spin-only ${\rm SWAP}$ yield a situation where the spin qubit
from DQD $k$ and the valley qubit from DQD $m$ are together in DQD 2, where the
universal two-qubit gate can be applied \cite{rohling_burkard}.
If two valley qubits or two spin qubits need to be involved in the two-qubit gate, the spin and the valley qubits
can be interchanged when they are in DQD 2.
The average number of additional ${\rm SWAP}$ gates needed for transferring qubits to the DQD at
one end of the register before the desired gates can be applied is on the order of $n^2$ \cite{Supplemental}.
The register Fig.~\ref{fig:register} (b) contains a spin-only DQD for every spin-valley DQD.
This allows for single-qubit operations without transferring qubits through the whole register.
Two-qubit gates between arbitrary qubits can be performed as spins can be moved within the entire register
by spin-only ${\rm SWAP}$ gates.
It is crucial to achieve high-fidelity
spin-only and spin-valley SWAP gates.
Errors in those operations can lead to leakage as spin and valley could leave
the singlet-triplet subspaces.

\textit{Materials.}---We now focus on the materials which may provide the necessary
properties for realizing our QC scheme.
A natural choice seems to be a hybrid structure of one material without and
one with a valley degeneracy where the hopping  parameters $t$ and
$\varphi$ across the interface can be controlled.

The sixfold valley degeneracy of bulk silicon
is typically split off by a (001) interface
\cite{RevModPhys.54.437,PhysRevB.61.2961,herring,PhysRev.101.944,PhysRevB.34.5621,people_bean,PhysRevB.48.14276,schaeffler};
the two lower valleys of interest, denoted $\ket{z}$ and $\ket{\overline z}$,
are coupled by
valley-orbit interaction,
described by the complex matrix element $V_{\rm VO}$ \cite{PhysRevB.80.081305}.
In experiment, a valley splitting $2|V_{\rm VO}|$ in lateral silicon
quantum dots of 0.1-1~meV has been reported
\cite{xiao:032103,borselli:123118,borselli:063109,lim,simmons,PhysRevB.86.115319,yang2013,PhysRevLett.111.046801}.
Electrical tunability in the range from 0.3 to 0.8~meV
was demonstrated \cite{yang2013}.
Calculations show that
while $\arg(V_{\rm VO})$ is only slightly influenced by an applied electric field \cite{PhysRevB.86.035321},
it depends
on the conduction band offset \cite{PhysRevB.84.155320};
thus,
a structure with an alternating top layer material, e.g., SiO$_2$ and SiGe
or Si$_{1{-}x}$Ge$_x$ with a varying $x$ could provide
the difference in $\arg(V_{\rm VO})$, i.e. $\varphi\neq0$, which is needed in our scheme between dots I and II
and between III and IV, while $\arg(V_{\rm VO})$ should be the same in dot II and IV.
Whereas the band offset difference is higher between Si/SiO$_2$ and Si/SiGe,
charge traps at the Si/SiO$_2$ interface can be a source for noise \cite{APL95_0723102};
nevertheless, spin blockade was demonstrated in Si/SiO$_2$ structures
with a reduced density of traps \cite{SciRep1_110}.
Furthermore, one requires a large valley splitting in dots I and III, which
results in a situation where effectively only one valley state
participates in the dynamics.

Graphene provides a twofold valley degeneracy and should allow for a
two-dimensional array of quantum dots.
Theory shows that the valley state is affected by a magnetic field
perpendicular to the graphene plane \cite{PhysRevB.79.085407}
and in graphene nanoribbons also by the boundary conditions \cite{trauzettel}.

\textit{Dual hybrid register.}---Interchanging the roles of
spin and valley yields DQDs with spin and valley DOF
and with valley DOF only.
A strong local magnetic field could yield conditions where only the lowest spin states
in this DQD have to be taken into account;
the spin Zeeman splitting has to be large compared to the exchange
interaction, e.g., 0.01~meV in Ref.~\cite{petta}.
To achieve a different phase between these spin states and the spins of the energy
eigenstates in the spin-valley dots, the direction of the
magnetic field has to be different for different dots.
Therefore, this alternative approach seems to have less stringent requirements from the material
point of view (phase of valley states can be the same), but would require a 
field gradient of several Tesla on a nanometer scale.

\textit{Conclusion.}---In conclusion, we have shown that combining spin and valley singlet-triplet
qubits allows for a new hybrid spin-valley QC scheme.
Necessary conditions are control over the Zeeman splitting for spin and valley
as well as over the phase of the valley states,
the realization of high-fidelity ${\rm SWAP}$ operations,
and long enough valley coherence times.
The concept relies crucially on the controlled coexistence of spin and valley qubits
allowing for universal QC based on the electrically tunable exchange
interaction.

\textit{Acknowledgements.}---We thank the DFG for financial support under
Programs No. SPP 1285 and No. SFB 767.

\onecolumngrid

\newpage

\begin{center}\textbf{\large Supplemental Material}\end{center}

\section{A. Schrieffer-Wolff transformation}
\stepcounter{section}
The Schrieffer-Wolff transformation for our Hamiltonian from the main text,
\begin{equation}
 \tilde H =e^{-S}He^S\approx H_0+H_V+\frac{[H_T,S]}{2}=\left(\begin{array}{cc}H_{\rm eff} & 0\\
                                                                                  0       & H_{\rm as}
                                                             \end{array}\right),
\tag{A1}
\end{equation}
is done similar to the situation without the valley degree of freedom [S1].
The block $H_{\rm eff}$ describes the physics in the low-energy subspace, which
has approximately (1,1) charge configuration when the detuning $\varepsilon$ is
close to zero.
We can perform the transformation for the spin singlet subspace and for the spin
triplet subspaces separately.
The projection on the three-dimensional subspace including one spin triplet,
is spanned e.g. by the basis
$\{\ket{\uparrow,{\uparrow\!+}}=\hat c_{1\uparrow}^\dagger\hat c_{2\uparrow+}^\dagger\ket{0},\ket{\uparrow,{\uparrow\!-}}\hat c_{1\uparrow}^\dagger\hat c_{2\uparrow-}^\dagger\ket{0},\ket{0,{\uparrow\!+}{\uparrow\!-}}=\hat c_{2\uparrow+}\hat c_{2\uparrow-}\ket{0}\}$
for the $T_+$ triplet and described by the Hamiltonian,
\begin{equation}
 H_{\rm Triplet} = \left(\begin{array}{ccc}h    & 0     & -t_-\\
                                           0    & -h     & t_+\\
                                         -t_-^* & t_+^* & U-\varepsilon
                         \end{array}\right).
\tag{A2}
\end{equation}
For the spin singlet subspace we have in the basis
$\{(\ket{\uparrow,\downarrow\!+}-\ket{\downarrow,\uparrow\!+})/\sqrt{2},
(\ket{\uparrow,\downarrow\!-}-\ket{\downarrow,\uparrow\!-})/\sqrt{2},
(\ket{0,{\uparrow\!+}{\downarrow\!-}}+\ket{0,{\uparrow\!-}{\downarrow\!+}})/\sqrt{2},
\ket{\uparrow\downarrow,0},
\ket{0,{\uparrow\!+}{\downarrow\!+}},
\ket{0,{\uparrow\!-}{\downarrow\!-}}\}$,
the Hamiltonian
\begin{equation}
 H_{\rm Singlet} = \left(\begin{array}{cccccc}
                          h     &  0             &      t_-       & \sqrt{2}t_+^* & \sqrt{2}t_+      &   0\\
                          0     &  h             &      t_+       & \sqrt{2}t_-^* &      0           &   \sqrt{2}t_-^*  \\
                         t_-^*  & t_+^*          &  U-\varepsilon &      0        &      0           &   0  \\
                   \sqrt{2}t_+  & \sqrt{2}t_-    &     0          & U+\varepsilon &      0           &   0   \\
                   \sqrt{2}t_+^*&  0             &     0          &      0        & U-\varepsilon+2h &   0    \\
                         0      & \sqrt{2}t_-^*  &     0          &      0        &      0           & U-\varepsilon-2h
                         \end{array}\right).
\tag{A3}
\end{equation}
The anti-Hermitian matrix $S$ should obey $[H_0+H_V,S]=-H_T$, which is fulfilled if
the blocks corresponding to the singlet and triplet subspaces are given by
\begin{equation}
 S_{\rm Triplet}=\left(\begin{array}{ccc}
                        0                   &        0                       & \frac{t_-}{h-U+\varepsilon} \\
                        0                   &        0                       & \frac{-t_+}{-h-U+\varepsilon}\\
             \frac{-t_-^*}{h-U+\varepsilon} & \frac{t_+^*}{-h-U+\varepsilon} &           0
                       \end{array}\right)
\tag{A4}
\end{equation}
and
\begin{equation}
 S_{\rm Singlet} = \left(\begin{array}{cccccc}
                          0            &          0                          & \frac{-t_-}{h-U+\varepsilon} & \frac{\sqrt{2}t_+^*}{-h+U+\varepsilon} & \frac{-\sqrt{2}t_+}{-h-U+\varepsilon} & 0\\
                          0            &          0                          & \frac{-t_+}{-h-U+\varepsilon}& \frac{\sqrt{2}t_-^*}{h+U+\varepsilon}  &          0                            &\frac{-\sqrt{2}t_-}{h-U+\varepsilon}\\
         \frac{t_-^*}{h-U+\varepsilon} & \frac{t_+^*}{-h-U+\varepsilon}      & 0 & 0 & 0 & 0\\
 \frac{-\sqrt{2}t_+}{-h+U+\varepsilon} & \frac{-\sqrt{2}t_-}{h+U+\varepsilon}& 0 & 0 & 0 & 0\\
 \frac{\sqrt{2}t_+^*}{-h-U+\varepsilon}& 0 & 0 & 0 & 0 & 0\\
                          0            & \frac{\sqrt{2}t_-^*}{h-U+\varepsilon} &0 & 0 & 0 & 0
                         \end{array}\right).
\tag{A5}
\end{equation}
This leads to the effective Hamiltonian given in Eq.~(1) of the main text.

\section{B. Time evolution with the effective Hamiltonian}
\stepcounter{section}
\setcounter{Bild}{1}
The time evolution according to the effective Hamiltonian,
\begin{equation}
 H_{\rm eff} =  \left[(A\cos\varphi{+}B)\mathds{1} + (A{+}B\cos\varphi)\sigma_z + B\sin\varphi\,\sigma_y\right]P_S
               + \tilde h\sigma_z
                +C\left[(1-\cos\varphi)\mathds{1}-\sin\varphi\,\sigma_y\right],
\tag{B1}               
\end{equation}
can be considered separately for the spin singlet subspace and for the
three spin triplet subspaces, where the latter are identical to each other.
We denote the two-dimensional effective Hamiltonians for the singlet and
one of the triplet subspaces by
\begin{equation}
 H_{\rm eff}^S = (A\cos\varphi{+}B{+}C(1-\cos\varphi))\mathds{1} + (\tilde h{+}A{+}B\cos\varphi)\sigma_z + (B-C)\sin\varphi\,\sigma_y
\tag{B2}
\end{equation}
and
\begin{equation}
 H_{\rm eff}^T = \tilde h\sigma_z + C\left[(1-\cos\varphi)\mathds{1}-\sin\varphi\,\sigma_y\right].
\tag{B3}
\end{equation}
By introducing
\begin{equation}
 \theta_S=\sqrt{(\tilde h{+}A{+}B\cos\varphi)^2+(B-C)^2\sin^2\varphi}, \hspace{1cm} \vec n_S=\frac{(B-C)\sin\varphi}{\theta_S}\vec e_y + \frac{\tilde h{+}A{+}B\cos\varphi}{\theta_S}\vec e_z,
\tag{B4}
\end{equation}
\begin{equation}
 \theta_T=\sqrt{\tilde h^2+C^2\sin^2\varphi}, \hspace{1cm} \vec n_T=\frac{-C\sin\varphi}{\theta_T}\vec e_y + \frac{\tilde h}{\theta_T}\vec e_z,
\tag{B5}
\end{equation}
we find for the corresponding time evolution operators
\begin{equation}
 U_S(\tau) = e^{-iH^S_{\rm eff}\tau/\hbar} = e^{-i(A\cos\varphi{+}B{+}C(1-\cos\varphi))\tau/\hbar}(\cos(\theta_S\tau/\hbar)\mathds{1}-i\sin(\theta_S\tau/\hbar)\vec n_S\cdot \boldsymbol{\sigma}),
\tag{B6}
\end{equation}
\begin{equation}
 U_T(\tau) = e^{-iH^T_{\rm eff}\tau/\hbar} = e^{-iC(1-\cos\varphi)\tau/\hbar}(\cos(\theta_T\tau/\hbar)\mathds{1}-i\sin(\theta_T\tau/\hbar)\vec n_T\cdot \boldsymbol{\sigma}).
\tag{B7}
\end{equation}
In a basis with spin singlet and triplet states, say
$\{\ket{T_++},\ket{T_+-},\ket{T_0+},\ket{T_0-},\ket{T_-+},\ket{T_--},\ket{S+},\ket{S-}\}$
where $\pm$ denotes the valley state of the electron in the right dot
and $T_+,T_-,T_0,S$ the spin state, e.g., $\ket{T_++}=\ket{\uparrow,\uparrow\!+}$,
the time evolution operator $U_{\rm eff}$ is block diagonal,
\begin{equation}
 U_{\rm eff} (\tau) = \left(\begin{array}{cccc} U_T(\tau) &     0     &     0     &   0\\
                                                   0      & U_T(\tau) &     0     &   0\\
                                                   0      &     0     & U_T(\tau) &   0\\
                                                   0      &     0     &     0     & U_S(\tau)
                            \end{array}\right).
\tag{B8}
\end{equation}
 The average fidelity of this operation with respect to a spin SWAP gate,
 ${\rm SWAP}={\rm SWAP}^\dagger=\operatorname{diag}(1,1,1,1,1,1,-1,-1)$
in our basis, maximized over a unimportant $z$ rotation in valley space
is given by (see [S2] for general formula)
\begin{equation}
F = \max_\alpha\frac{8+|\operatorname{Tr}(e^{i\alpha\sigma_z}U_{\rm eff}(\tau){\rm SWAP})|^2}{8(8+1)}
  = \max_\alpha\frac{8+|3\operatorname{Tr}(e^{i\alpha\sigma_z}U_T(\tau))-\operatorname{Tr}(e^{i\alpha\sigma_z}U_S(\tau))|^2}{72}
 \tag{B9}
\end{equation}
and $T(\tau):=\frac{1}{4}|3\operatorname{Tr}(e^{i\alpha\sigma_z}U_T(\tau))-\operatorname{Tr}(e^{i\alpha\sigma_z}U_S(\tau))|^2$ is according to (B6) and (B7) given by
\begin{equation}
T(\tau) = |3(\cos\alpha\cos(\theta_T\tau/\hbar)-\sin\alpha\sin(\theta_T\tau/\hbar)n_{Tz}) -e^{i(A\cos\varphi+B)\tau/\hbar}(\cos\alpha\cos(\theta_S\tau/\hbar)-\sin\alpha\sin(\theta_S\tau/\hbar)n_{Sz})|^2.
 \tag{B10}
\end{equation}
For $|A|,|B|,|C|\ll|\tilde h|$ we find the first order approximations
\begin{equation}
 \theta_T \approx \tilde h,\hspace{1cm} \theta_S \approx \tilde h + A + B\cos\varphi,\hspace{1cm}\text{and} \hspace{1cm}n_{Tz}\approx n_{Sz}\approx 1
\tag{B11}
\end{equation}
leading to
\begin{equation}
\begin{split}
 T(\tau) = & [3\cos(\alpha{+}\tilde h\tfrac{\tau}{\hbar}) - \cos((A\cos\varphi{+}B)\tfrac{\tau}{\hbar})\cos(\alpha{+}(\tilde h{+}A{+}B\cos\varphi)\tfrac{\tau}{\hbar})]^2\\
           &   + \sin^2((A\cos\varphi{+}B)\tfrac{\tau}{\hbar})\cos^2(\alpha{+}(\tilde h {+}A {+} B\cos\varphi)\tfrac{\tau}{\hbar})\\
         = & 9\cos^2(\alpha{+}\tilde h \tfrac{\tau}{\hbar}) + \cos^2(\alpha{+}(\tilde h{+}A{+}B\cos\varphi)\tfrac{\tau}{\hbar}) - 6\cos(\alpha{+}\tilde h\tfrac{\tau}{\hbar})\cos(\alpha{+}(\tilde h{+}A{+}B\cos\varphi)\tfrac{\tau}{\hbar})\cos((A\cos\varphi{+}B)\tfrac{\tau}{\hbar})].
\end{split}
\tag{B12}
\end{equation}
To find the value of $\alpha$ which gives the maximal average fidelity $F$, we calculate
\begin{equation}
 0\stackrel{!}{=}\frac{\partial T(\tau)}{\partial\alpha}
               = -9\sin(2\alpha{+}2\tilde h\tfrac{\tau}{\hbar}) - \sin(2\alpha{+}2(\tilde h{+}A{+}B\cos\varphi)\tfrac{\tau}{\hbar})+6\cos((A\cos\varphi{+}B)\tfrac{\tau}{\hbar})\sin(2\alpha{+}(2\tilde h{+}A{+}B\cos\varphi)\tfrac{\tau}{\hbar}),
\tag{B13}
\end{equation}
which is solved by
\begin{equation}
 \alpha = -\tilde h\frac{\tau}{\hbar} +\frac{1}{2} \arctan\frac{6\cos((A\cos\varphi+B)\frac{\tau}{\hbar})\sin((A+B\cos\varphi)\frac{\tau}{\hbar}) - \sin(2(A+B\cos\varphi)\frac{\tau}{\hbar})}{9+\cos(2(A+B\cos\varphi)\frac{\tau}{\hbar})-6\cos((A\cos\varphi+B)\frac{\tau}{\hbar})\cos((A+B\cos\varphi)\frac{\tau}{\hbar})},
\tag{B14}
\end{equation}
where the dominant term is $-\tilde h \frac{\tau}{\hbar}$.
The maximization with respect to $\tau$ is done numerically by using
minimization of $1-F$,
for the Figures 2 and 3 of the main text we calculate $F$ according
to Eq.~(B9) with the (approximate) maximization value for $\alpha$ from (B14).

\newpage
\section{C. Fidelity around first, second, third, and fourth local maximum}
\begin{figure}[hb]
\includegraphics[width=1\columnwidth]{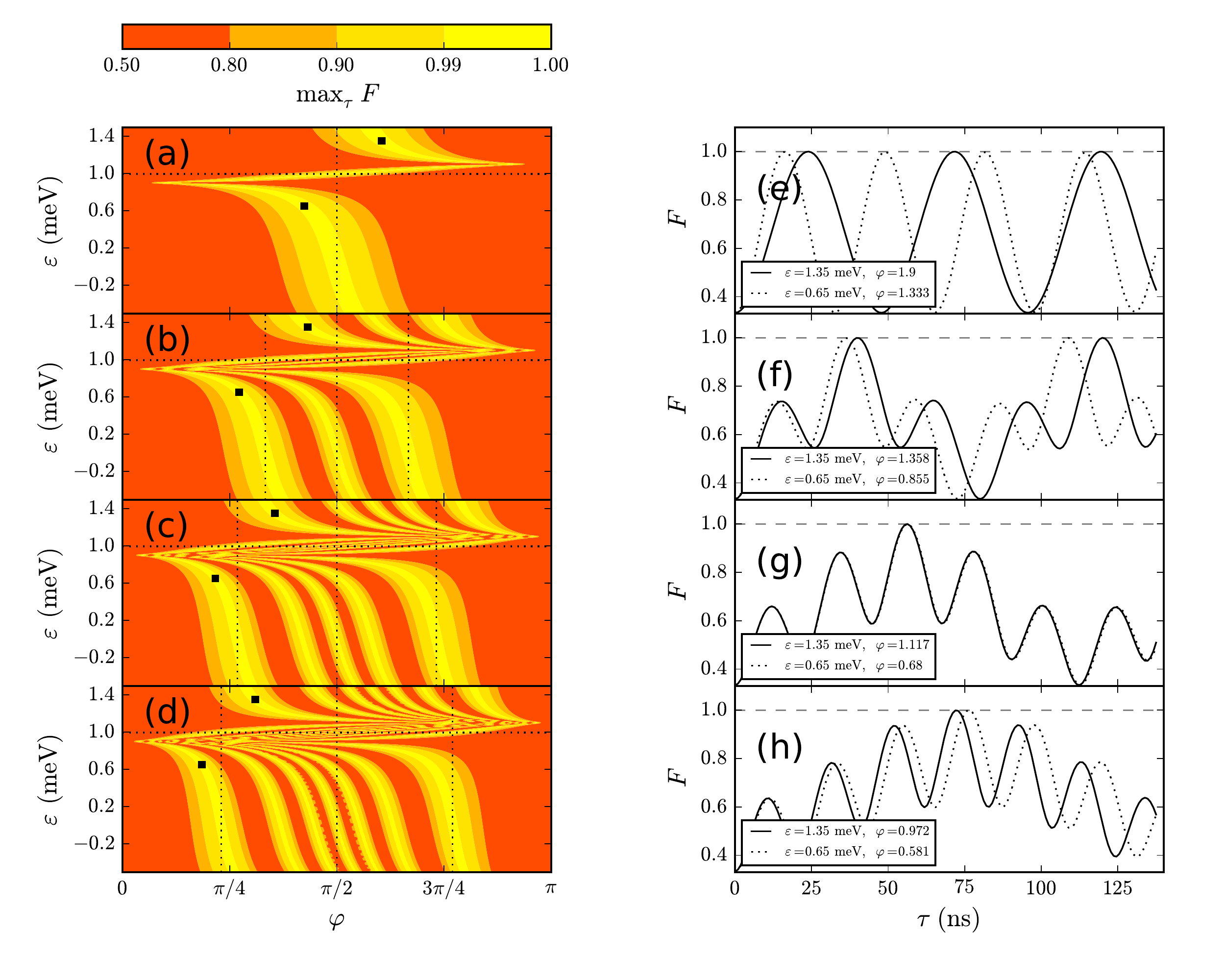} 
\caption{Fidelity $F$ for a broader range of
the detuning $\varepsilon$ and the phase in the hopping matrix elements, $\varphi$;
the parameters $U$, $h$, and $t$ are the same as in
Fig.~2 of the main text.
(a,b) $F$ maximized over the gate time $\tau$ (scale bar).  
We consider the first (a), second (b), third (c), and fourth (d) local maximum of $F$.
The black squares indicates the parameters for the time-dependent plots in (e)-(h).
The horizontal dotted lines in (a)-(d) represent $\varepsilon=\sqrt{U^2{-}h^2}$, and
the vertical dotted lines belong to values of $\varphi$ being $\frac{\pi}{2}$ (a-d);
$\frac{\pi}{3}$, $\frac{2\pi}{3}$ (b); $\arccos\frac{2}{3}$, $\pi-\arccos\frac{2}{3}$ (c);
$\arccos\frac{3}{4}$, $\pi-\arccos\frac{3}{4}$ (d).
Note that the Schrieffer-Wolff transformation breaks down close to $\varepsilon=U\pm h$.}
\label{fig:extended}
\end{figure}

\stepcounter{section}
\setcounter{Bild}{1}
In order to find more phases in the hopping matrix element, $\varphi$,
where the spin-only SWAP gate can be realized, we calculate $F$ for a broader
range of $\varepsilon$ and $\varphi$ than it was done for Fig.~2.
We consider the first, second, third, and fourth local maximum of the
fidelity as a function of time, $F(\tau)$.
The results are shown in Fig.~\ref{fig:extended}.

\stepcounter{Bild}

\begin{figure}[ht]
\includegraphics[width=.5\columnwidth]{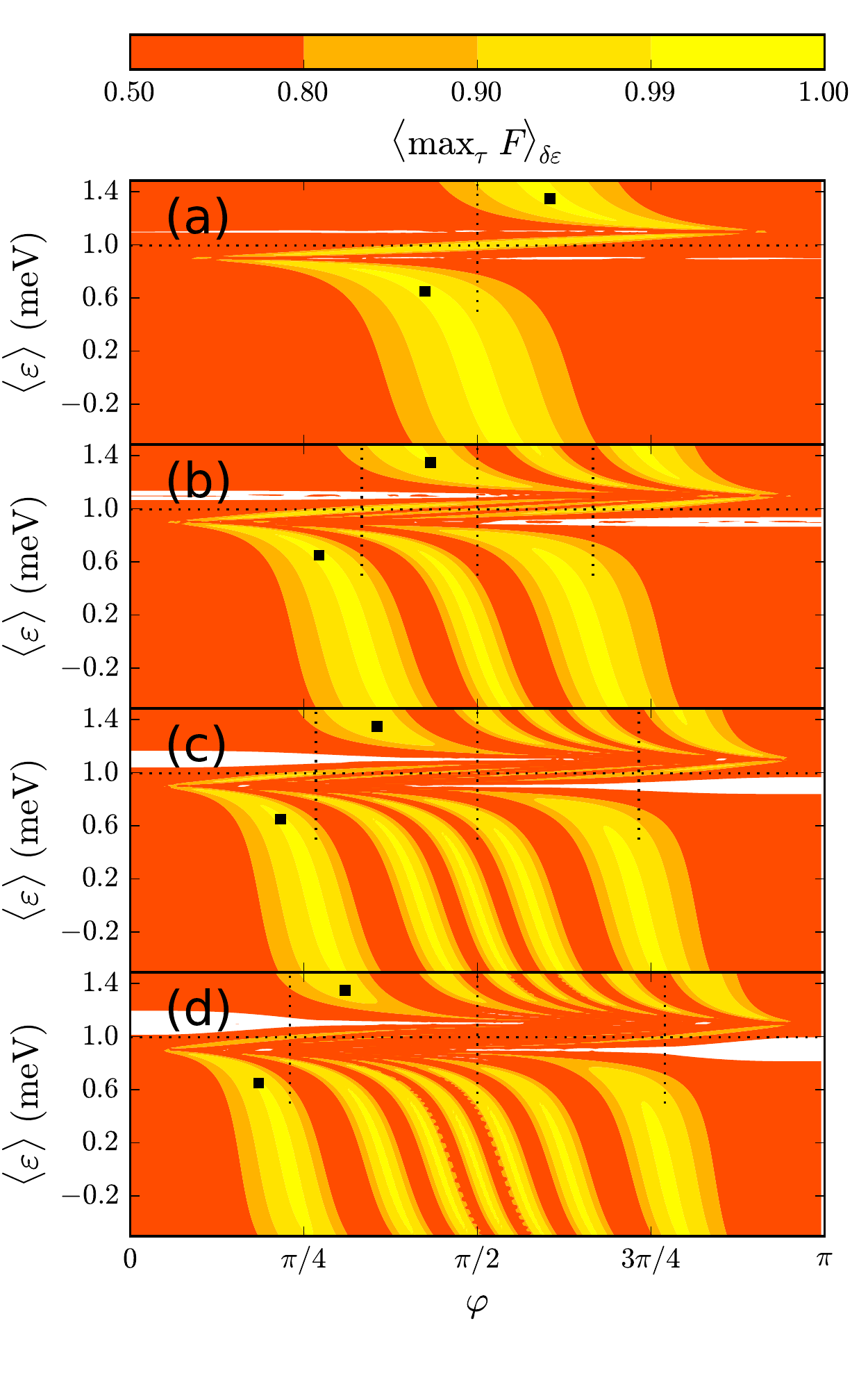} 
\caption{Fidelity in dependence of mean value of detuning $\langle\varepsilon\rangle$
and valley phase difference $\varphi$ for an average taken over a Gaussian
distribution of $\varepsilon$ with standard deviation of
$\sqrt{\langle\varepsilon^2\rangle-\langle\varepsilon\rangle^2}=8\mu$eV;
this value was found in [S3] for a DQD in GaAs.
While some small areas with $F>0.99$ from Fig.~\ref{fig:extended} are washed out
the broader stripes of high fidelity ``survive''.}
\label{fig:withnoise_extended}
\end{figure}

\newpage
By averaging the Fidelity $F$ over a Gaussian distribution of $\varepsilon$
while keeping the time from a maximization for a fixed value of $\varepsilon$
we can include the effect of quasi-static noise in our model.
Fig.~\ref{fig:withnoise_extended} shows that there are some regions of high fidelity which
seem to be quite robust against quasi-static noise.
\stepcounter{Bild}

\begin{figure}[h]
 \includegraphics[width=.5\textwidth]{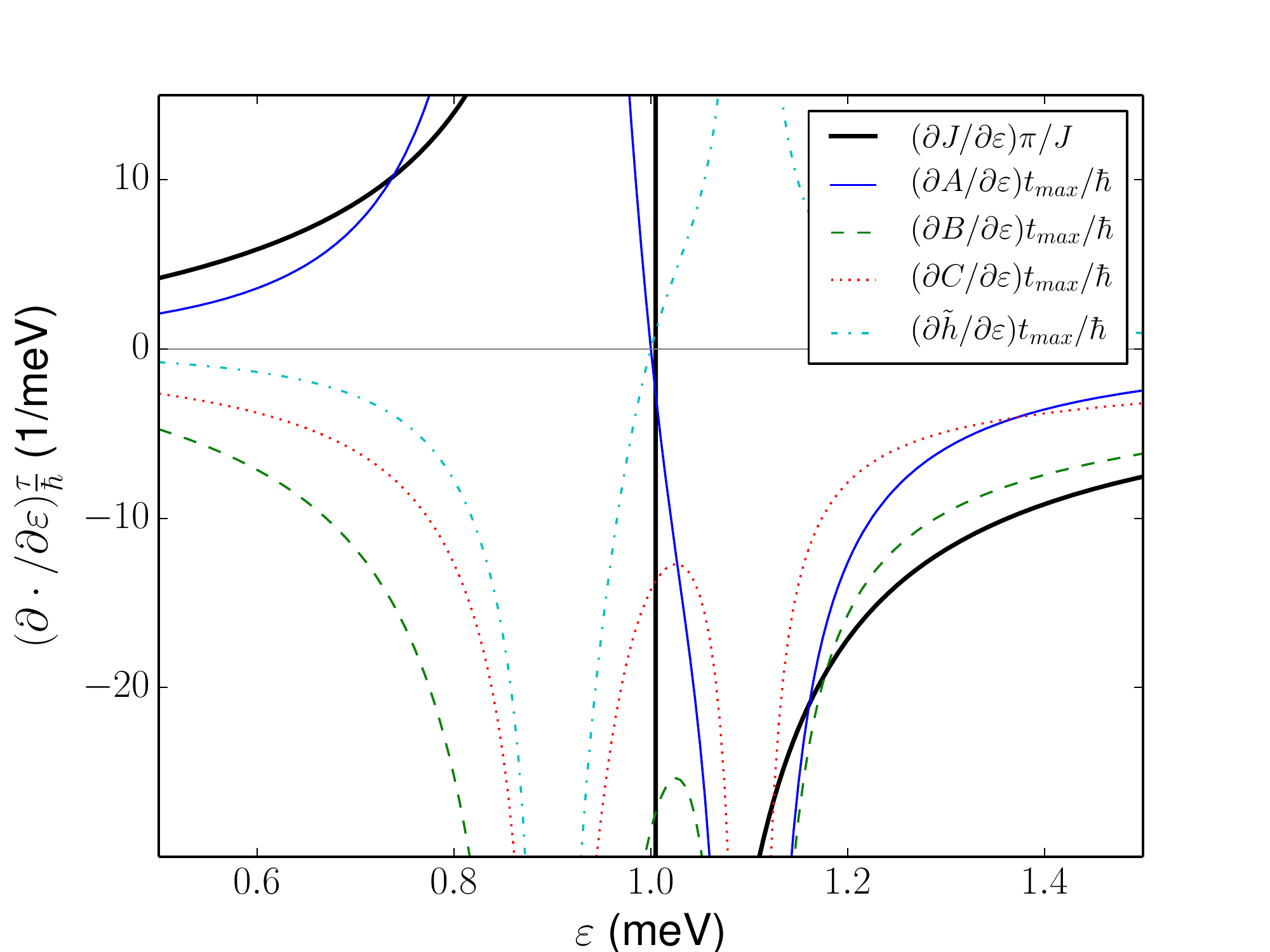}
\caption{Derivatives of the quantities in the Hamiltonian in
Eq. (1) of the main text (Eq. (B1) of this Supplemental Material)
in comparison to $J=4t^2U/(U^2-\varepsilon^2)$ for the spin-only exchange interaction.
In order to have the comparison to the time needed for a ${\rm SWAP_{spin}}$ operation,
we multiplied the derivatives with the time, when the fidelity of this quantum gate reaches
first local maximum.
The valley phase difference is given by $\varphi=1.333$.}
\label{fig:derivatives}
\end{figure}

We want to estimate the sensitivity of our ${\rm SWAP_{spin}}$ operation to electrical noise
in comparison to the valley-free case.
If no valley degeneracy is present the exchange interaction Hamiltonian is $-JP_S$ where $P_S$
is the projector on the spin singlet and, using the same assumptions as for our model,
$J=4t^2U/(U^2-\varepsilon^2)$.
The sensitivity to electrical noise is in first order given by $\partial J/\partial\varepsilon$ [S4].
Therefore we compare $\partial J/\partial\varepsilon$ to the derivatives of the quantities
$A$, $B$, $C$, and $\tilde h$ in $H_{\rm eff}$, see Eq. (1) of the main text and Eq. (B1) of
this Supplemental Material.
As the influence of the noise has to be considered relative to the gate time we plot in
Fig.~\ref{fig:derivatives} the product
of the derivative and the time when the first local maximum of the fidelity is reached.
In Fig.~\ref{fig:derivatives} the hopping phase is $\varphi=1.333$.
For the valley-free case this time is $\pi\hbar/J$.
This is of course only a rough estimation, but at $\varepsilon\approx0.65\,{\rm meV}$,
which is the value where a very high fidelity is reached, the absolute values for the
quantities from our $H_{\rm eff}$ are in the same order of magnitude as
$(\partial J/\partial \varepsilon)\pi/J$.
Thus we expect the noise sensitivity of our exchange-based interaction to be similar to the
situation without valley degeneracy.

It should be mentioned that for a fixed phase difference $\varphi$ we need to choose
$\varepsilon$ such that the fidelity $F$ can be high
whereas in the case without valley, $\varepsilon$ can be tuned in order to achieve low noise influence
only worsen the gating time.

Furthermore noise can in principle also affect the gate via the
tunneling parameter $t$, as it is done in Ref.~[S5] we neglect this effect here.

\section{D. Explicit sequences for quantum gates}
\stepcounter{section}
\setcounter{Bild}{1}
In this section we give explicit sequences which show how single- and two-qubit
gates can be applied in the two registers shown in Fig.~1 of the main text.
In following we consider first the register Fig.~1 (a) and after that Fig.~1 (b).
We introduce the notation
${\rm SWAP}_{sv}(m,m-1)$
for the $\rm SWAP_{\rm spin} \otimes \rm SWAP_{\rm valley}$ gate
applied at the upper and the lower rows of the register between the dots at position $m$ and $m-1$.
This gate is provided for dots with spin and valley degeneracy directly by exchange interaction [S6],
i.e., in Fig.~1 (a) they can be applied for $m\le3$.
Furthermore we denote the spin-only SWAP gate between the dots at position
$m$ and $m-1$, applied again in the upper and the lower line of the register,
by ${\rm SWAP}_{s}(m,m-1)$.
This gate can be performed for the register
Fig.~1 (a) for $m=2$ and for
Fig.~1 (b) for any $m=2,3,4,\ldots$
The single qubit gates of the valley qubit in the $k$-th double quantum dot (DQD) in
Fig.~1 (a) are realized by the  four steps below:
\begin{enumerate}
\item Apply ${\rm SWAP}_{sv}(k,k-1), {\rm SWAP}_{sv}(k-1,k-2),\ldots,{\rm SWAP}_{sv}(3,2)$.
\item Apply ${\rm SWAP}_{s}(2,1)$.
\item Perform the single-qubit operation in DQD 2.
\item Repeat the second step and then the first in inverse order to bring the qubits back to position $k$.
\end{enumerate}
The exchange interaction and a gradient in the valley splitting provides full
control over two axis in Bloch sphere of the valley qubit and thus allow for the third step.
Single-qubit operations with the $S$-$T_0$ spin triplet are done the same way using exchange
and a gradient in the spin Zeeman field of DQD 1 within step 3 of the given procedure.
For a two-qubit gate between the valley qubits of DQD $k$ and $m$ with $m>k$ the steps are
as follows:
\begin{enumerate}
\item Apply ${\rm SWAP}_{sv}(k,k-1), {\rm SWAP}_{sv}(k-1,k-2),\ldots,{\rm SWAP}_{sv}(3,2)$ and
${\rm SWAP}_{sv}(m,m-1), {\rm SWAP}_{sv}(m-1,m-2),\ldots,{\rm SWAP}_{sv}(4,3)$.
\item Apply a SWAP gate between the spin and the valley qubit residing
  in DQD $k$.
This is possible as any unitary operation is feasible in this subsystem.
\item Apply ${\rm SWAP}_{s}(2,1)$.
\item Apply ${\rm SWAP}_{sv}(3,2)$.
\item Apply ${\rm SWAP}_{s}(2,1)$.
\item Apply the  two-qubit gate between the spin and the valley qubit in
DQD 2, which are the former valley qubits of DQDs $k$ and $m$.
\item Return the qubits to their former position by reversing steps 1 to 5.
\end{enumerate}
For a two-qubit gate between a spin and a valley qubit, step 2 is not necessary and
between two spin qubits, step 2 exchanges the spin and valley qubits from the $m$-th
instead of the $k$-th DQD. 

\begin{figure}[ht]
 \begin{minipage}{0.45\textwidth}
  \includegraphics[width=7cm]{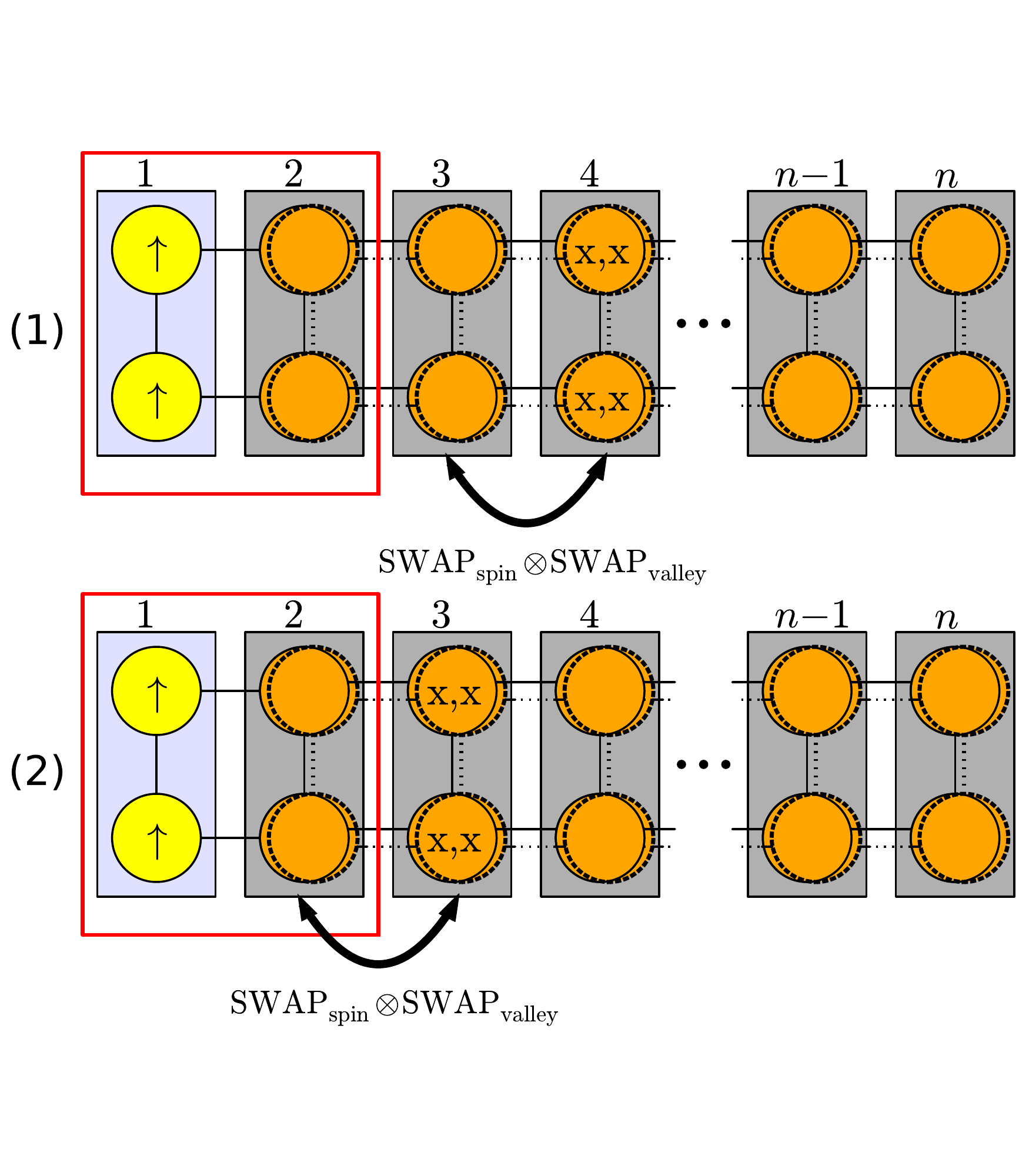}
 \end{minipage}
 \begin{minipage}{0.45\textwidth}
  \includegraphics[width=7cm]{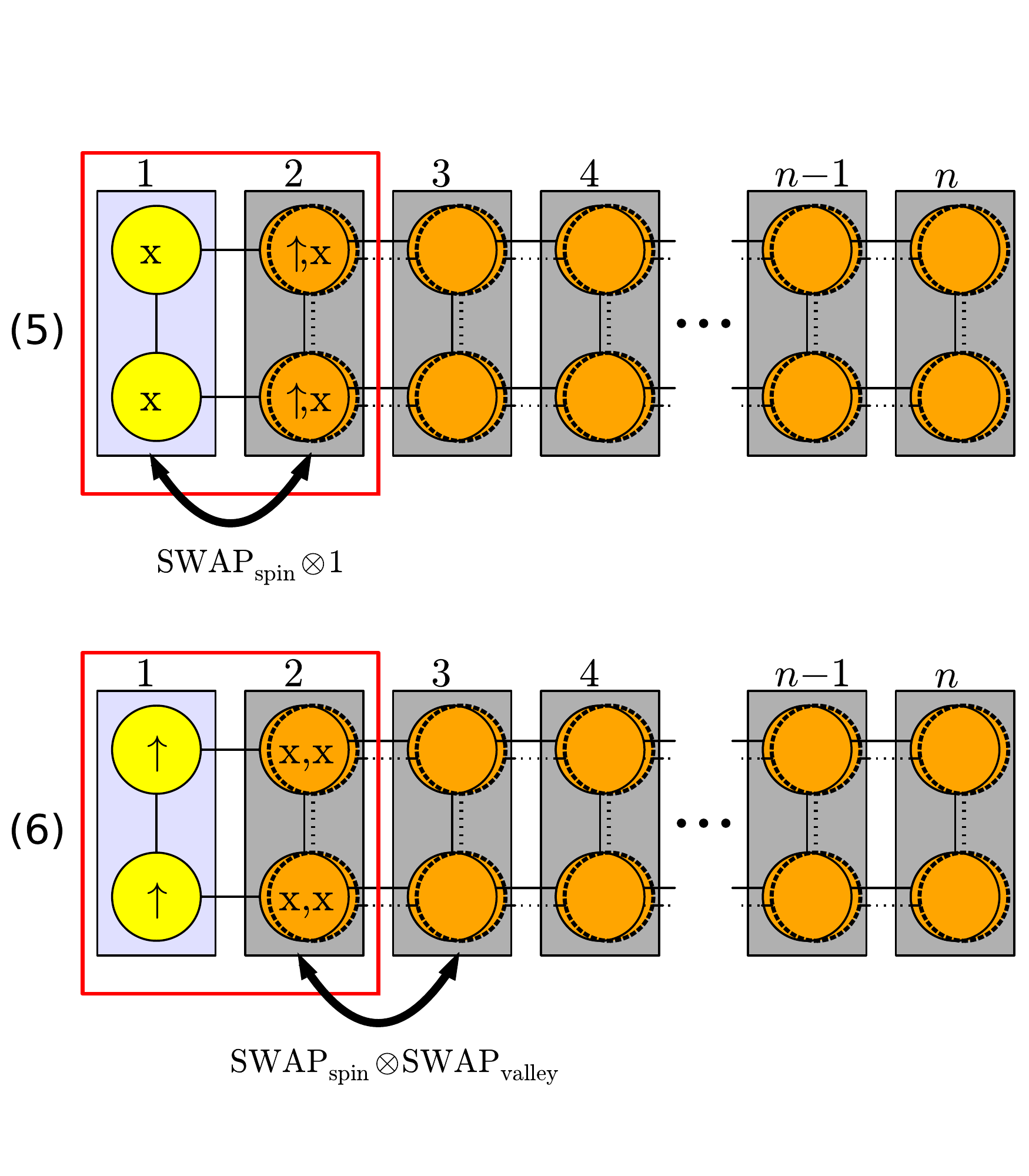}
 \end{minipage}
 \begin{minipage}{0.45\textwidth}
  \includegraphics[width=7cm]{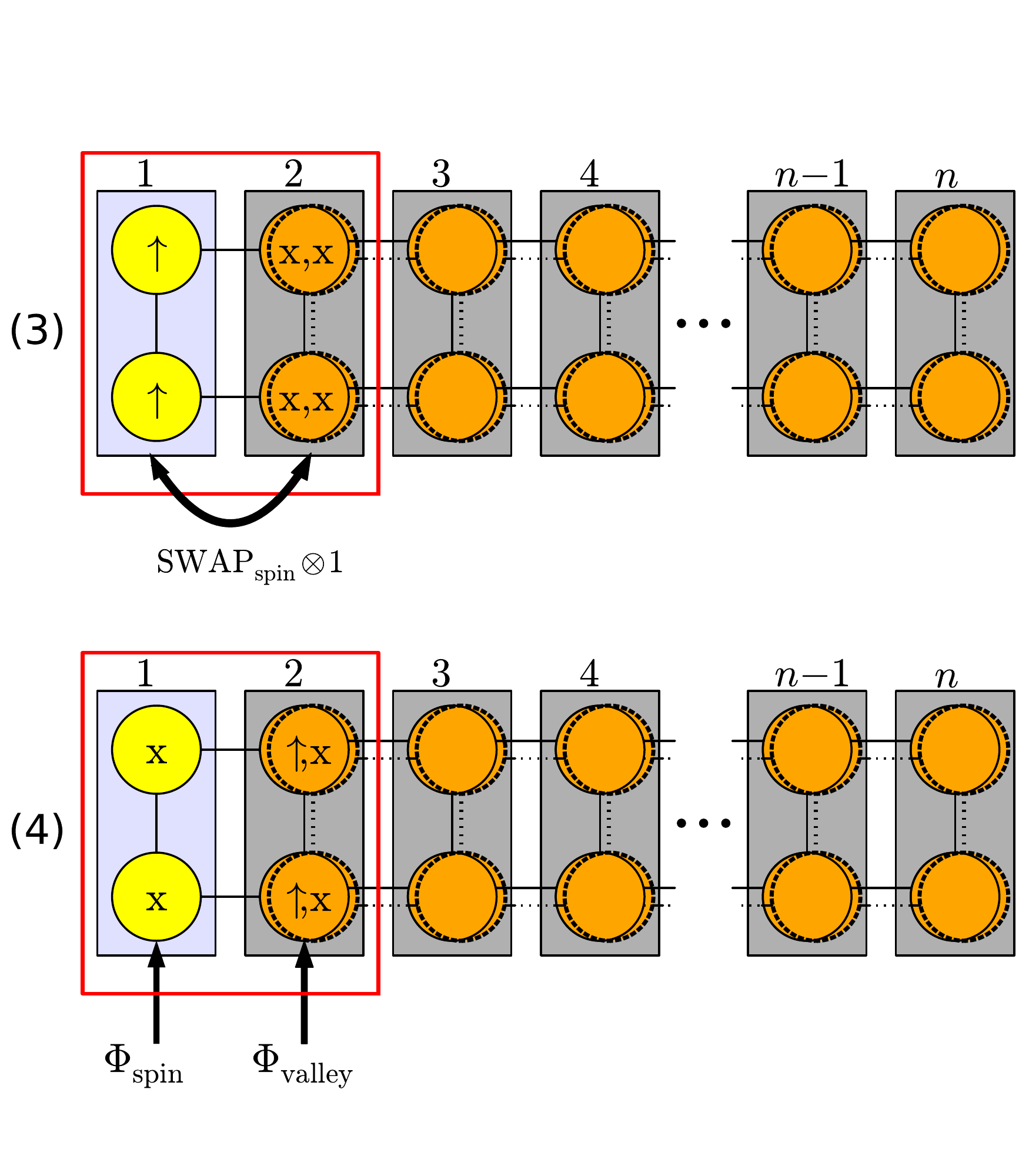}
 \end{minipage}
 \begin{minipage}{0.45\textwidth}
  \includegraphics[width=7cm]{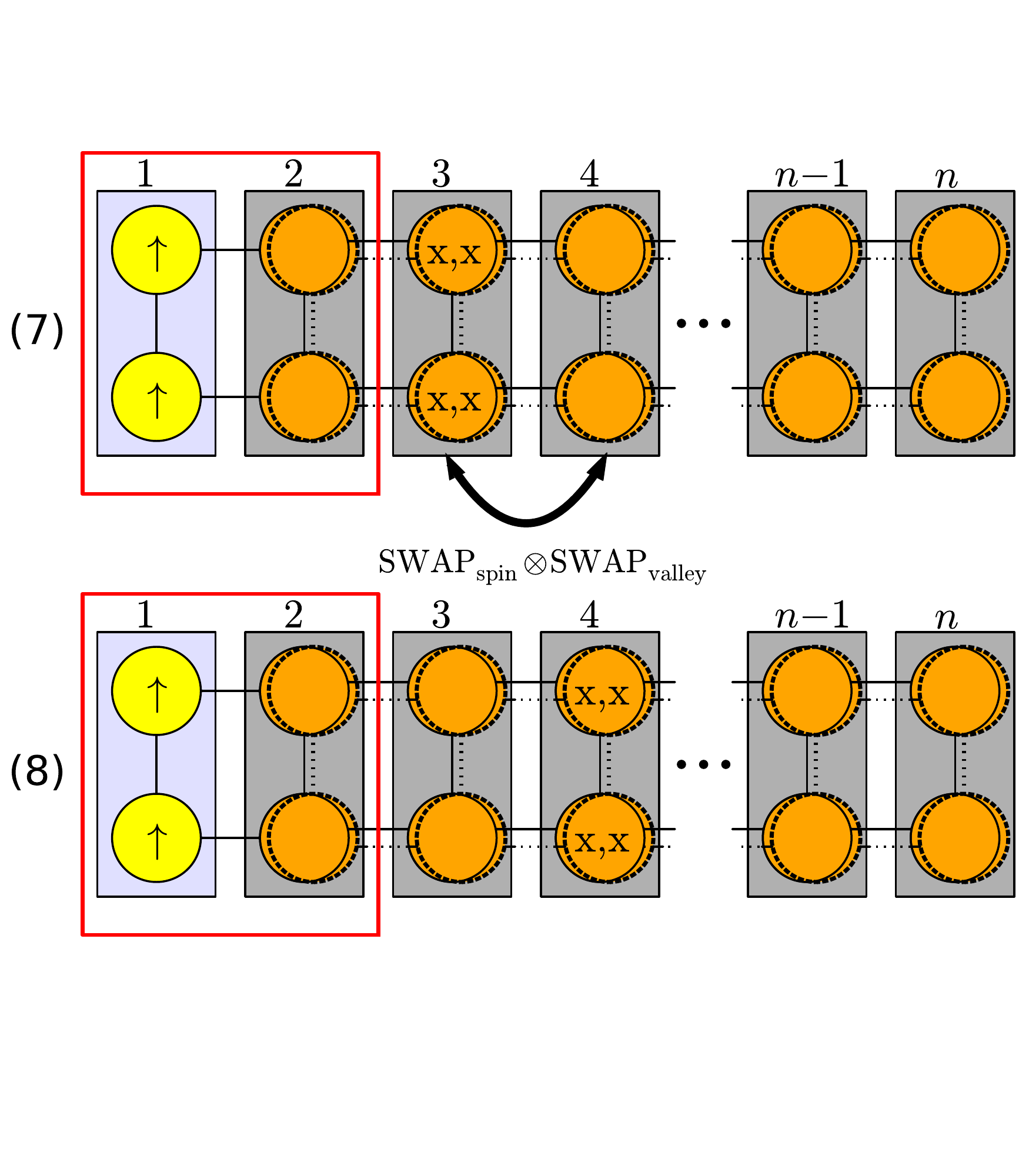}
 \end{minipage}
\caption{Performing single-qubit gates on the spin and the valley qubits in DQD 4 for register of Fig.~1 (a) of the main text.
First, the qubits are moved to DQD 2 by ${\rm SWAP_{spin}}\otimes{\rm SWAP_{valley}}$ in step (1) and (2),
then spin and valley qubit are separated in step (3),
in step (4) the single-qubit gates are applied and finally by reversing (1-3) in steps (5-7)
the qubits are moved back to their original position (8).
In total $2\cdot4-1=7$ gates are needed, note that all ${\rm SWAP}$ operations are done
in parallel in the upper and the lower rows of the register in parallel and
the single-qubit operations  in step (4) can be performed in parallel in DQD 1 and DQD 2.}
\label{fig:explicit}
\end{figure}

We proved that universal quantum computing is possible
in Fig.~1 (a), but should also ask how efficient our
scheme is, in particular, whether it scales polynomially or not. 
To answer this
question, we count the number of gates needed for a single-qubit operation each
on the whole $2n-2$ qubits.
We remember for a single gate on a qubit in the $k$-th DQD we need each
$2(k-2)$ $\rm SWAP_{\rm spin} \otimes \rm SWAP_{\rm valley}$ gates for swapping the qubits
to position 2 and back and two spin-only ${\rm SWAP}$ gates;
all those gates are applied on the upper and the lower line of the register in parallel (so only counted once).
Then for the qubits in the $k$-th DQD  we need $2k-1$ quantum gates
($2k-4$ $\rm SWAP_{\rm spin} \otimes \rm SWAP_{\rm valley}$ gates,
two spin-only ${\rm SWAP}$,
and the single qubit gates,
which we only count as one because they can be applied in parallel on the spin $S$-$T_0$ qubit in DQD 1
and on the valley $S$-$T_0$ qubit in DQD 2), see Fig. \ref{fig:explicit} for $k=4$.
For applying single-qubit gates on each qubit in the register this means
a total number of $\sum_{k=2}^n (2k-1)=O(n^2)$ needed quantum gates.

\newpage
Now we consider Fig.~1 (b) of the main text and prove universality
alike the register in Fig.~1 (a). The single-qubit gates are directly given by the alternating
structure of the register.
For the two-qubit gates between two valley qubits in DQD $k$ and $m>k$ the steps
are as follows:
\begin{enumerate}
\item Swap the valley qubits with the spin qubit in DQD $m$.
\item Apply ${\rm SWAP}_{s}(m,m-1), {\rm SWAP}_{s}(m-1,m-2),\ldots, {\rm SWAP}_{s}(k+1,k)$.
\item Apply the two-qubit gate between the spin and the valley in DQD $m$.
\item Return the qubits to their former position by reversing step 2 and 1.
\end{enumerate} 
For a two-qubit gate between a spin and a valley qubit, step 1 is not necessary and between
two spin qubits step 2 interchanges the spin and the valley from the $m$-th instead of the
$k$-th qubit.

\begin{itemize}
 \item[{[S1]}] G.~Burkard and A.~Imamoglu, \href{\doibase 10.1103/PhysRevB.74.041307} {Phys. Rev. B \textbf{74}, 041307 (2006)}.

\item[{[S2]}]
  L.~H. Pedersen, N.~M. M{\o}ller, and K.~M{\o}lmer,
  \href{\doibase 10.1016/j.physleta.2007.02.069}
  {Phys. Lett. A \textbf{367}, 47 (2007)}.

\item[{[S3]}]
  O.~E. Dial, M.~D. Shulman, S.~P. Harvey, H. Bluhm, V. Umansky, and A. Yacoby,
  \href{\doibase 10.1103/PhysRevLett.110.146804}
  {Phys. Rev. Lett. \textbf{110}, 146804 (2013)}

\item[{[S4]}]
  M.~P. Wardrop and A.~C. Doherty,
  \href{http://link.aps.org/doi/10.1103/PhysRevB.90.045418}
  {Phys. Rev. B \textbf{90}, 045418 (2014)}.

\item[{[S5]}]
  D. Culcer, X. Hu, and S. Das~Sarma
  \href {http://dx.doi.org/10.1063/1.3194778}
  {Appl. Phys. Lett. \textbf{95}, 073102 (2009)}.

\item[{[S6]}]
  N.~Rohling and G.~Burkard, \href{http://stacks.iop.org/1367-2630/14/i=8/a=083008}
  {New J. Phys. \textbf{14}, 083008 (2012)}.
\end{itemize}


\begin{thebibliography}{78}%
\makeatletter
\providecommand \@ifxundefined [1]{%
 \@ifx{#1\undefined}
}%
\providecommand \@ifnum [1]{%
 \ifnum #1\expandafter \@firstoftwo
 \else \expandafter \@secondoftwo
 \fi
}%
\providecommand \@ifx [1]{%
 \ifx #1\expandafter \@firstoftwo
 \else \expandafter \@secondoftwo
 \fi
}%
\providecommand \natexlab [1]{#1}%
\providecommand \enquote  [1]{``#1''}%
\providecommand \bibnamefont  [1]{#1}%
\providecommand \bibfnamefont [1]{#1}%
\providecommand \citenamefont [1]{#1}%
\providecommand \href@noop [0]{\@secondoftwo}%
\providecommand \href [0]{\begingroup \@sanitize@url \@href}%
\providecommand \@href[1]{\@@startlink{#1}\@@href}%
\providecommand \@@href[1]{\endgroup#1\@@endlink}%
\providecommand \@sanitize@url [0]{\catcode `\\12\catcode `\$12\catcode
  `\&12\catcode `\#12\catcode `\^12\catcode `\_12\catcode `\%12\relax}%
\providecommand \@@startlink[1]{}%
\providecommand \@@endlink[0]{}%
\providecommand \url  [0]{\begingroup\@sanitize@url \@url }%
\providecommand \@url [1]{\endgroup\@href {#1}{\urlprefix }}%
\providecommand \urlprefix  [0]{URL }%
\providecommand \Eprint [0]{\href }%
\providecommand \doibase [0]{http://dx.doi.org/}%
\providecommand \selectlanguage [0]{\@gobble}%
\providecommand \bibinfo  [0]{\@secondoftwo}%
\providecommand \bibfield  [0]{\@secondoftwo}%
\providecommand \translation [1]{[#1]}%
\providecommand \bibitemNoStop [0]{.\EOS\space}%
\providecommand \EOS [0]{\spacefactor3000\relax}%
\providecommand \BibitemShut  [1]{\csname bibitem#1\endcsname}%
\let\auto@bib@innerbib\@empty
\bibitem{LoDi1998}
  \bibfield  {author} {\bibinfo {author} {\bibfnamefont {D.}~\bibnamefont
  {Loss}}\ and\ \bibinfo {author} {\bibfnamefont {D.~P.}\ \bibnamefont
  {DiVincenzo}},\ }\href {\doibase 10.1103/PhysRevA.57.120} {\bibfield
  {journal} {\bibinfo  {journal} {Phys. Rev. A}\ }\textbf {\bibinfo {volume}
  {57}},\ \bibinfo {pages} {120} (\bibinfo {year} {1998})}.

\bibitem{kloeffel_loss_review_2013}%
  {For a recent review including both approaches, see }\bibfield  {author} {\bibinfo {author} {\bibfnamefont {C.}~\bibnamefont
  {Kloef\-fel}}\ and\ \bibinfo {author} {\bibfnamefont {D.}~\bibnamefont
  {Loss}},\ }\href {\doibase 10.1146/annurev-conmatphys-030212-184248}
  {\bibfield  {journal} {\bibinfo  {journal} {Annu. Rev. of Condens. Matter
  Phys.}\ }\textbf {\bibinfo {volume} {4}},\ \bibinfo {pages} {51} (\bibinfo
  {year} {2013})}.

\bibitem [{\citenamefont {Petta}\ \emph {et~al.}(2005)\citenamefont {Petta},
  \citenamefont {Johnson}, \citenamefont {Taylor}, \citenamefont {Laird},
  \citenamefont {Yacoby}, \citenamefont {Lukin}, \citenamefont {Marcus},
  \citenamefont {Hanson},\ and\ \citenamefont {Gossard}}]{petta}%
  \bibfield  {author} {\bibinfo {author} {\bibfnamefont {J.~R.}\ \bibnamefont
  {Petta}}, \bibinfo {author} {\bibfnamefont {A.~C.}\ \bibnamefont {Johnson}},
  \bibinfo {author} {\bibfnamefont {J.~M.}\ \bibnamefont {Taylor}}, \bibinfo
  {author} {\bibfnamefont {E.~A.}\ \bibnamefont {Laird}}, \bibinfo {author}
  {\bibfnamefont {A.}~\bibnamefont {Yacoby}}, \bibinfo {author} {\bibfnamefont
  {M.~D.}\ \bibnamefont {Lukin}}, \bibinfo {author} {\bibfnamefont {C.~M.}\
  \bibnamefont {Marcus}}, \bibinfo {author} {\bibfnamefont {M.~P.}\
  \bibnamefont {Hanson}}, \ and\ \bibinfo {author} {\bibfnamefont {A.~C.}\
  \bibnamefont {Gossard}},\ }\href {\doibase 10.1126/science.1116955}
  {\bibfield  {journal} {\bibinfo  {journal} {Science}\ }\textbf {\bibinfo
  {volume} {309}},\ \bibinfo {pages} {2180} (\bibinfo {year}
  {2005})}.

\bibitem [{\citenamefont {Foletti}\ \emph {et~al.}(2009)\citenamefont
  {Foletti}, \citenamefont {Bluhm}, \citenamefont {Mahalu}, \citenamefont
  {Umansky},\ and\ \citenamefont {Yacoby}}]{foletti2009}%
  \bibfield  {author} {\bibinfo {author} {\bibfnamefont {S.}~\bibnamefont
  {Foletti}}, \bibinfo {author} {\bibfnamefont {H.}~\bibnamefont {Bluhm}},
  \bibinfo {author} {\bibfnamefont {D.}~\bibnamefont {Mahalu}}, \bibinfo
  {author} {\bibfnamefont {V.}~\bibnamefont {Umansky}}, \ and\ \bibinfo
  {author} {\bibfnamefont {A.}~\bibnamefont {Yacoby}},\ }\href
  {http://dx.doi.org/10.1038/nphys1424} {\bibfield  {journal} {\bibinfo
  {journal} {Nat. Phys.}\ }\textbf {\bibinfo {volume} {5}},\ \bibinfo
  {pages} {903} (\bibinfo {year} {2009})}.

\bibitem [{\citenamefont {Levy}(2002)}]{Levy2002}%
  \bibfield  {author} {\bibinfo {author} {\bibfnamefont {J.}~\bibnamefont
  {Levy}},\ }\href {\doibase 10.1103/PhysRevLett.89.147902} {\bibfield
  {journal} {\bibinfo  {journal} {Phys. Rev. Lett.}\ }\textbf {\bibinfo
  {volume} {89}},\ \bibinfo {pages} {147902} (\bibinfo {year}
  {2002})}.

\bibitem [{\citenamefont {Benjamin}(2001)}]{Benjamin_pra2001}%
  \bibfield  {author} {\bibinfo {author} {\bibfnamefont {S.~C.}\ \bibnamefont
  {Benjamin}},\ }\href {\doibase 10.1103/PhysRevA.64.054303} {\bibfield
  {journal} {\bibinfo  {journal} {Phys. Rev. A}\ }\textbf {\bibinfo {volume}
  {64}},\ \bibinfo {pages} {054303} (\bibinfo {year} {2001})}.

\bibitem [{\citenamefont {Klinovaja}\ \emph {et~al.}(2012)\citenamefont
  {Klinovaja}, \citenamefont {Stepanenko}, \citenamefont {Halperin},\ and\
  \citenamefont {Loss}}]{Klinovaja_prb2012}%
  \bibfield  {author} {\bibinfo {author} {\bibfnamefont {J.}~\bibnamefont
  {Klinovaja}}, \bibinfo {author} {\bibfnamefont {D.}~\bibnamefont
  {Stepanenko}}, \bibinfo {author} {\bibfnamefont {B.~I.}\ \bibnamefont
  {Halperin}}, \ and\ \bibinfo {author} {\bibfnamefont {D.}~\bibnamefont
  {Loss}},\ }\href {\doibase 10.1103/PhysRevB.86.085423} {\bibfield  {journal}
  {\bibinfo  {journal} {Phys. Rev. B}\ }\textbf {\bibinfo {volume} {86}},\
  \bibinfo {pages} {085423} (\bibinfo {year} {2012})}.

\bibitem{PhysRevB.86.205306}%
  \bibfield  {author} {\bibinfo {author} {\bibfnamefont {R.}~\bibnamefont
  {Li}}, \bibinfo {author} {\bibfnamefont {X.}~\bibnamefont {Hu}}, \ and\
  \bibinfo {author} {\bibfnamefont {J.~Q.}\ \bibnamefont {You}},\ }\href
  {\doibase 10.1103/PhysRevB.86.205306} {\bibfield  {journal} {\bibinfo
  {journal} {Phys. Rev. B}\ }\textbf {\bibinfo {volume} {86}},\ \bibinfo
  {pages} {205306} (\bibinfo {year} {2012})}.

\bibitem{Taylor_natphys}
  \bibfield  {author} {\bibinfo {author} {\bibfnamefont {J.~M.}\ \bibnamefont
  {Taylor}}, \bibinfo {author} {\bibfnamefont {H.-A.}\ \bibnamefont {Engel}},
  \bibinfo {author} {\bibfnamefont {W.}~\bibnamefont {Dur}}, \bibinfo {author}
  {\bibfnamefont {A.}~\bibnamefont {Yacoby}}, \bibinfo {author} {\bibfnamefont
  {C.~M.}\ \bibnamefont {Marcus}}, \bibinfo {author} {\bibfnamefont
  {P.}~\bibnamefont {Zoller}}, \ and\ \bibinfo {author} {\bibfnamefont {M.~D.}\
  \bibnamefont {Lukin}},\ }\href {\doibase 10.1038/nphys174} {\bibfield
  {journal} {\bibinfo  {journal} {Nat. Phys.}\ }\textbf {\bibinfo {volume}
  {1}},\ \bibinfo {pages} {177} (\bibinfo {year} {2005})}.

\bibitem{Stepanenko2007}%
  \bibfield  {author} {\bibinfo {author} {\bibfnamefont {D.}~\bibnamefont
  {Stepanenko}}\ and\ \bibinfo {author} {\bibfnamefont {G.}~\bibnamefont
  {Burkard}},\ }\href {\doibase 10.1103/PhysRevB.75.085324} {\bibfield
  {journal} {\bibinfo  {journal} {Phys. Rev. B}\ }\textbf {\bibinfo {volume}
  {75}},\ \bibinfo {pages} {085324} (\bibinfo {year} {2007})}.

\bibitem{PhysRevB.84.155329}%
  \bibfield  {author} {\bibinfo {author} {\bibfnamefont {G.}~\bibnamefont
  {Ramon}},\ }\href {\doibase 10.1103/PhysRevB.84.155329} {\bibfield  {journal}
  {\bibinfo  {journal} {Phys. Rev. B}\ }\textbf {\bibinfo {volume} {84}},\
  \bibinfo {pages} {155329} (\bibinfo {year} {2011})}.

\bibitem{PhysRevLett.107.030506}%
  \bibfield  {author} {\bibinfo {author} {\bibfnamefont {I.}~\bibnamefont {van
  Weperen}}, \bibinfo {author} {\bibfnamefont {B.~D.}\ \bibnamefont
  {Armstrong}}, \bibinfo {author} {\bibfnamefont {E.~A.}\ \bibnamefont
  {Laird}}, \bibinfo {author} {\bibfnamefont {J.}~\bibnamefont {Medford}},
  \bibinfo {author} {\bibfnamefont {C.~M.}\ \bibnamefont {Marcus}}, \bibinfo
  {author} {\bibfnamefont {M.~P.}\ \bibnamefont {Hanson}}, \ and\ \bibinfo
  {author} {\bibfnamefont {A.~C.}\ \bibnamefont {Gossard}},\ }\href {\doibase
  10.1103/PhysRevLett.107.030506} {\bibfield  {journal} {\bibinfo  {journal}
  {Phys. Rev. Lett.}\ }\textbf {\bibinfo {volume} {107}},\ \bibinfo {pages}
  {030506} (\bibinfo {year} {2011})}.

\bibitem{Shulman}%
  \bibfield  {author} {\bibinfo {author} {\bibfnamefont {M.~D.}\ \bibnamefont
  {Shulman}}, \bibinfo {author} {\bibfnamefont {O.~E.}\ \bibnamefont {Dial}},
  \bibinfo {author} {\bibfnamefont {S.~P.}\ \bibnamefont {Harvey}}, \bibinfo
  {author} {\bibfnamefont {H.}~\bibnamefont {Bluhm}}, \bibinfo {author}
  {\bibfnamefont {V.}~\bibnamefont {Umansky}}, \ and\ \bibinfo {author}
  {\bibfnamefont {A.}~\bibnamefont {Yacoby}},\ }\href {\doibase
  10.1126/science.1217692} {\bibfield  {journal} {\bibinfo  {journal}
  {Science}\ }\textbf {\bibinfo {volume} {336}},\ \bibinfo {pages} {202}
  (\bibinfo {year} {2012})}.

\bibitem{rohling_burkard}%
  \bibfield  {author} {\bibinfo {author} {\bibfnamefont {N.}~\bibnamefont
  {Rohling}}\ and\ \bibinfo {author} {\bibfnamefont {G.}~\bibnamefont
  {Burkard}},\ }\href {http://stacks.iop.org/1367-2630/14/i=8/a=083008}
  {\bibfield  {journal} {\bibinfo  {journal} {New J. Phys.}\ }\textbf
  {\bibinfo {volume} {14}},\ \bibinfo {pages} {083008} (\bibinfo {year}
  {2012})}.  

\bibitem{zwanenburg}%
  {For a recent review see }
  \bibfield  {author} {\bibinfo {author} {\bibfnamefont {F.~A.}\ \bibnamefont
  {Zwanenburg}}, \bibinfo {author} {\bibfnamefont {A.~S.}\ \bibnamefont
  {Dzurak}}, \bibinfo {author} {\bibfnamefont {A.}~\bibnamefont {Morello}},
  \bibinfo {author} {\bibfnamefont {M.~Y.}\ \bibnamefont {Simmons}}, \bibinfo
  {author} {\bibfnamefont {L.~C.~L.}\ \bibnamefont {Hollenberg}}, \bibinfo
  {author} {\bibfnamefont {G.}~\bibnamefont {Klimeck}}, \bibinfo {author}
  {\bibfnamefont {S.}~\bibnamefont {Rogge}}, \bibinfo {author} {\bibfnamefont
  {S.~N.}\ \bibnamefont {Coppersmith}}, \ and\ \bibinfo {author} {\bibfnamefont
  {M.~A.}\ \bibnamefont {Eriksson}},\ }\href {\doibase
  10.1103/RevModPhys.85.961} {\bibfield  {journal} {\bibinfo  {journal} {Rev.
  Mod. Phys.}\ }\textbf {\bibinfo {volume} {85}},\ \bibinfo {pages} {961}
  (\bibinfo {year} {2013})}.

\bibitem{castro_neto}%
  \bibfield  {author} {\bibinfo {author} {\bibfnamefont {A.~H.}\ \bibnamefont
  {Castro~Neto}}, \bibinfo {author} {\bibfnamefont {F.}~\bibnamefont {Guinea}},
  \bibinfo {author} {\bibfnamefont {N.~M.~R.}\ \bibnamefont {Peres}}, \bibinfo
  {author} {\bibfnamefont {K.~S.}\ \bibnamefont {Novoselov}}, \ and\ \bibinfo
  {author} {\bibfnamefont {A.~K.}\ \bibnamefont {Geim}},\ }\href {\doibase
  10.1103/RevModPhys.81.109} {\bibfield  {journal} {\bibinfo  {journal} {Rev.
  Mod. Phys.}\ }\textbf {\bibinfo {volume} {81}},\ \bibinfo {pages} {109}
  (\bibinfo {year} {2009})}.

\bibitem{kuemmeth}%
  \bibfield  {author} {\bibinfo {author} {\bibfnamefont {F.}~\bibnamefont
  {Kuemmeth}}, \bibinfo {author} {\bibfnamefont {S.}~\bibnamefont {Ilani}},
  \bibinfo {author} {\bibfnamefont {D.~C.}\ \bibnamefont {Ralph}}, \ and\
  \bibinfo {author} {\bibfnamefont {P.~L.}\ \bibnamefont {McEuen}},\
  }\href@noop {} {\bibfield  {journal} {\bibinfo  {journal} {Nature (London)}\ }\textbf
  {\bibinfo {volume} {452}},\ \bibinfo {pages} {448} (\bibinfo {year}
  {2008})}.

\bibitem{shayegan}%
  \bibfield  {author} {\bibinfo {author} {\bibfnamefont {M.}~\bibnamefont
  {Shayegan}}, \bibinfo {author} {\bibfnamefont {E.~P.}\ \bibnamefont
  {De~Poortere}}, \bibinfo {author} {\bibfnamefont {O.}~\bibnamefont
  {Gunawan}}, \bibinfo {author} {\bibfnamefont {Y.~P.}\ \bibnamefont
  {Shkolnikov}}, \bibinfo {author} {\bibfnamefont {E.}~\bibnamefont {Tutuc}}, \
  and\ \bibinfo {author} {\bibfnamefont {K.}~\bibnamefont {Vakili}},\ }\href
  {\doibase 10.1002/pssb.200642212} {\bibfield  {journal} {\bibinfo  {journal}
  {Phys. Status Solidi (b)}\ }\textbf {\bibinfo {volume} {243}},\ \bibinfo
  {pages} {3629} (\bibinfo {year} {2006})}.

\bibitem{xiao}%
  \bibfield  {author} {\bibinfo {author} {\bibfnamefont {D.}~\bibnamefont
  {Xiao}}, \bibinfo {author} {\bibfnamefont {G.-B.}\ \bibnamefont {Liu}},
  \bibinfo {author} {\bibfnamefont {W.}~\bibnamefont {Feng}}, \bibinfo {author}
  {\bibfnamefont {X.}~\bibnamefont {Xu}}, \ and\ \bibinfo {author}
  {\bibfnamefont {W.}~\bibnamefont {Yao}},\ }\href {\doibase
  10.1103/PhysRevLett.108.196802} {\bibfield  {journal} {\bibinfo  {journal}
  {Phys. Rev. Lett.}\ }\textbf {\bibinfo {volume} {108}},\ \bibinfo {pages}
  {196802} (\bibinfo {year} {2012})}.

\bibitem{kormanyos}%
  \bibfield  {author} {\bibinfo {author} {\bibfnamefont {A.}~\bibnamefont
  {Korm\'anyos}}, \bibinfo {author} {\bibfnamefont {V.}~\bibnamefont
  {Z\'olyomi}}, \bibinfo {author} {\bibfnamefont {N.~D.}\ \bibnamefont
  {Drummond}}, \ and\ \bibinfo {author} {\bibfnamefont {G.}~\bibnamefont
  {Burkard}},\ }\href {\doibase
  10.1103/PhysRevX.4.011034} {\bibfield  {journal} {\bibinfo  {journal}
  {Phys. Rev. X}\ }\textbf {\bibinfo {volume} {4}},\ \bibinfo {pages}
  {011034} (\bibinfo {year} {2014})}.

\bibitem{feher_gere}%
  \bibfield  {author} {\bibinfo {author} {\bibfnamefont {G.}~\bibnamefont
  {Feher}}\ and\ \bibinfo {author} {\bibfnamefont {E.~A.}\ \bibnamefont
  {Gere}},\ }\href {\doibase 10.1103/PhysRev.114.1245} {\bibfield  {journal}
  {\bibinfo  {journal} {Phys. Rev.}\ }\textbf {\bibinfo {volume} {114}},\
  \bibinfo {pages} {1245} (\bibinfo {year} {1959})}.

\bibitem{tyryshkin}%
  \bibfield  {author} {\bibinfo {author} {\bibfnamefont {A.~M.}\ \bibnamefont
  {Tyryshkin}}, \bibinfo {author} {\bibfnamefont {S.}~\bibnamefont {Tojo}},
  \bibinfo {author} {\bibfnamefont {J.~J.~L.}\ \bibnamefont {Morton}}, \bibinfo
  {author} {\bibfnamefont {H.}~\bibnamefont {Riemann}}, \bibinfo {author}
  {\bibfnamefont {N.~V.}\ \bibnamefont {Abrosimov}}, \bibinfo {author}
  {\bibfnamefont {P.}~\bibnamefont {Becker}}, \bibinfo {author} {\bibfnamefont
  {H.-J.}\ \bibnamefont {Pohl}}, \bibinfo {author} {\bibfnamefont
  {T.}~\bibnamefont {Schenkel}}, \bibinfo {author} {\bibfnamefont {M.~L.~W.}\
  \bibnamefont {Thewalt}}, \bibinfo {author} {\bibfnamefont {K.~M.}\
  \bibnamefont {Itoh}}, \ and\ \bibinfo {author} {\bibfnamefont {S.~A.}\
  \bibnamefont {Lyon}},\ }\href {\doibase 10.1038/nmat3182} {\bibfield
  {journal} {\bibinfo  {journal} {Nat. Mater.}\ }\textbf {\bibinfo
  {volume} {11}},\ \bibinfo {pages} {143} (\bibinfo {year} {2012})}.

\bibitem{boykin:115}%
  \bibfield  {author} {\bibinfo {author} {\bibfnamefont {T.~B.}\ \bibnamefont
  {Boykin}}, \bibinfo {author} {\bibfnamefont {G.}~\bibnamefont {Klimeck}},
  \bibinfo {author} {\bibfnamefont {M.~A.}\ \bibnamefont {Eriksson}}, \bibinfo
  {author} {\bibfnamefont {M.}~\bibnamefont {Friesen}}, \bibinfo {author}
  {\bibfnamefont {S.~N.}\ \bibnamefont {Coppersmith}}, \bibinfo {author}
  {\bibfnamefont {P.}~\bibnamefont {von Allmen}}, \bibinfo {author}
  {\bibfnamefont {F.}~\bibnamefont {Oyafuso}}, \ and\ \bibinfo {author}
  {\bibfnamefont {S.}~\bibnamefont {Lee}},\ }\href {\doibase 10.1063/1.1637718}
  {\bibfield  {journal} {\bibinfo  {journal} {Appl. Phys. Lett.}\
  }\textbf {\bibinfo {volume} {84}},\ \bibinfo {pages} {115} (\bibinfo {year}
  {2004})}.

\bibitem{nestoklon}%
  \bibfield  {author} {\bibinfo {author} {\bibfnamefont {M.~O.}\ \bibnamefont
  {Nestoklon}}, \bibinfo {author} {\bibfnamefont {L.~E.}\ \bibnamefont
  {Golub}}, \ and\ \bibinfo {author} {\bibfnamefont {E.~L.}\ \bibnamefont
  {Ivchenko}},\ }\href {\doibase 10.1103/PhysRevB.73.235334} {\bibfield
  {journal} {\bibinfo  {journal} {Phys. Rev. B}\ }\textbf {\bibinfo {volume}
  {73}},\ \bibinfo {pages} {235334} (\bibinfo {year} {2006})}.

\bibitem {chutia}%
  \bibfield  {author} {\bibinfo {author} {\bibfnamefont {S.}~\bibnamefont
  {Chutia}}, \bibinfo {author} {\bibfnamefont {S.~N.}\ \bibnamefont
  {Coppersmith}}, \ and\ \bibinfo {author} {\bibfnamefont {M.}~\bibnamefont
  {Friesen}},\ }\href {\doibase 10.1103/PhysRevB.77.193311} {\bibfield
  {journal} {\bibinfo  {journal} {Phys. Rev. B}\ }\textbf {\bibinfo {volume}
  {77}},\ \bibinfo {pages} {193311} (\bibinfo {year} {2008})}.

\bibitem{PhysRevB.80.081305}%
  \bibfield  {author} {\bibinfo {author} {\bibfnamefont {A.~L.}\ \bibnamefont
  {Saraiva}}, \bibinfo {author} {\bibfnamefont {M.~J.}\ \bibnamefont
  {Calder\'on}}, \bibinfo {author} {\bibfnamefont {X.}~\bibnamefont {Hu}},
  \bibinfo {author} {\bibfnamefont {S.}~\bibnamefont {Das~Sarma}}, \ and\
  \bibinfo {author} {\bibfnamefont {B.}~\bibnamefont {Koiller}},\ }\href
  {\doibase 10.1103/PhysRevB.80.081305} {\bibfield  {journal} {\bibinfo
  {journal} {Phys. Rev. B}\ }\textbf {\bibinfo {volume} {80}},\ \bibinfo
  {pages} {081305} (\bibinfo {year} {2009})}.

\bibitem{PhysRevB.82.155312}%
  \bibfield  {author} {\bibinfo {author} {\bibfnamefont {D.}~\bibnamefont
  {Culcer}}, \bibinfo {author} {\bibfnamefont {L.}~\bibnamefont
  {Cywi\ifmmode~\acute{n}\else \'{n}\fi{}ski}}, \bibinfo {author}
  {\bibfnamefont {Q.}~\bibnamefont {Li}}, \bibinfo {author} {\bibfnamefont
  {X.}~\bibnamefont {Hu}}, \ and\ \bibinfo {author} {\bibfnamefont
  {S.}~\bibnamefont {Das~Sarma}},\ }\href {\doibase 10.1103/PhysRevB.82.155312}
  {\bibfield  {journal} {\bibinfo  {journal} {Phys. Rev. B}\ }\textbf {\bibinfo
  {volume} {82}},\ \bibinfo {pages} {155312} (\bibinfo {year}
  {2010}{\natexlab{b}})}.

\bibitem{PhysRevB.82.245314}%
  \bibfield  {author} {\bibinfo {author} {\bibfnamefont {A.~L.}\ \bibnamefont
  {Saraiva}}, \bibinfo {author} {\bibfnamefont {B.}~\bibnamefont {Koiller}}, \
  and\ \bibinfo {author} {\bibfnamefont {M.}~\bibnamefont {Friesen}},\ }\href
  {\doibase 10.1103/PhysRevB.82.245314} {\bibfield  {journal} {\bibinfo
  {journal} {Phys. Rev. B}\ }\textbf {\bibinfo {volume} {82}},\ \bibinfo
  {pages} {245314} (\bibinfo {year} {2010})}.

\bibitem{PhysRevB.84.155320}%
  \bibfield  {author} {\bibinfo {author} {\bibfnamefont {A.~L.}\ \bibnamefont
  {Saraiva}}, \bibinfo {author} {\bibfnamefont {M.~J.}\ \bibnamefont
  {Calder\'on}}, \bibinfo {author} {\bibfnamefont {R.~B.}\ \bibnamefont
  {Capaz}}, \bibinfo {author} {\bibfnamefont {X.}~\bibnamefont {Hu}}, \bibinfo
  {author} {\bibfnamefont {S.}~\bibnamefont {Das~Sarma}}, \ and\ \bibinfo
  {author} {\bibfnamefont {B.}~\bibnamefont {Koiller}},\ }\href {\doibase
  10.1103/PhysRevB.84.155320} {\bibfield  {journal} {\bibinfo  {journal} {Phys.
  Rev. B}\ }\textbf {\bibinfo {volume} {84}},\ \bibinfo {pages} {155320}
  (\bibinfo {year} {2011})}.

\bibitem{baena}%
  \bibfield  {author} {\bibinfo {author} {\bibfnamefont {A.}~\bibnamefont
  {Baena}}, \bibinfo {author} {\bibfnamefont {A.~L.}\ \bibnamefont {Saraiva}},
  \bibinfo {author} {\bibfnamefont {B.}~\bibnamefont {Koiller}}, \ and\
  \bibinfo {author} {\bibfnamefont {M.~J.}\ \bibnamefont {Calder\'on}},\ }\href
  {\doibase 10.1103/PhysRevB.86.035317} {\bibfield  {journal} {\bibinfo
  {journal} {Phys. Rev. B}\ }\textbf {\bibinfo {volume} {86}},\ \bibinfo
  {pages} {035317} (\bibinfo {year} {2012})}.

\bibitem{drumm}%
  \bibfield  {author} {\bibinfo {author} {\bibfnamefont {D.~W.}\ \bibnamefont
  {Drumm}}, \bibinfo {author} {\bibfnamefont {A.}~\bibnamefont {Budi}},
  \bibinfo {author} {\bibfnamefont {M.~C.}\ \bibnamefont {Per}}, \bibinfo
  {author} {\bibfnamefont {S.~P.}\ \bibnamefont {Russo}}, \ and\ \bibinfo
  {author} {\bibfnamefont {L.~C.~L.}\ \bibnamefont {Hollenberg}},\ }\href
  {\doibase 10.1186/1556-276X-8-111} {\bibfield  {journal} {\bibinfo  {journal}
  {Nanoscale Res. Lett.}\ }\textbf {\bibinfo {volume} {8}},\ \bibinfo
  {pages} {111} (\bibinfo {year} {2013})}.

\bibitem {de_michelis}%
  \bibfield  {author} {\bibinfo {author} {\bibfnamefont {M.}\ \bibnamefont
  {De~Michielis}}, \bibinfo {author} {\bibfnamefont {E.}~\bibnamefont {Prati}},
  \bibinfo {author} {\bibfnamefont {M.}~\bibnamefont {Fanciulli}}, \bibinfo
  {author} {\bibfnamefont {G.}~\bibnamefont {Fiori}}, \ and\ \bibinfo {author}
  {\bibfnamefont {G.}~\bibnamefont {Iannaccone}},\ }\href {\doibase
  10.1143/APEX.5.124001} {\bibfield  {journal} {\bibinfo  {journal} {Appl.
  Phys. Express}\ }\textbf {\bibinfo {volume} {5}},\ \bibinfo {pages}
  {124001} (\bibinfo {year} {2012})}.

\bibitem{gamble2012}%
  \bibfield  {author} {\bibinfo {author} {\bibfnamefont {J.~K.}\ \bibnamefont
  {Gamble}}, \bibinfo {author} {\bibfnamefont {M.}~\bibnamefont {Friesen}},
  \bibinfo {author} {\bibfnamefont {S.~N.}\ \bibnamefont {Coppersmith}}, \ and\
  \bibinfo {author} {\bibfnamefont {X.}~\bibnamefont {Hu}},\ }\href {\doibase
  10.1103/PhysRevB.86.035302} {\bibfield  {journal} {\bibinfo  {journal} {Phys.
  Rev. B}\ }\textbf {\bibinfo {volume} {86}},\ \bibinfo {pages} {035302}
  (\bibinfo {year} {2012})}.

\bibitem{gamble2013}%
  \bibfield  {author} {\bibinfo {author} {\bibfnamefont {J.~K.}\ \bibnamefont
  {Gamble}}, \bibinfo {author} {\bibfnamefont {M.~A.}\ \bibnamefont
  {Eriksson}}, \bibinfo {author} {\bibfnamefont {S.~N.}\ \bibnamefont
  {Coppersmith}}, \ and\ \bibinfo {author} {\bibfnamefont {M.}~\bibnamefont
  {Friesen}},\ }\href {\doibase 10.1103/PhysRevB.88.035310} {\bibfield
  {journal} {\bibinfo  {journal} {Phys. Rev. B}\ }\textbf {\bibinfo {volume}
  {88}},\ \bibinfo {pages} {035310} (\bibinfo {year} {2013})}.

\bibitem{tahan_joynt_2013}%
  \bibfield  {author} {\bibinfo {author} {\bibfnamefont {C.}~\bibnamefont
  {Tahan}}\ and\ \bibinfo {author} {\bibfnamefont {R.}~\bibnamefont {Joynt}},\
  }\href {\doibase 10.1103/PhysRevB.89.075302} {\bibfield
  {journal} {\bibinfo  {journal} {Phys. Rev. B}\ }\textbf {\bibinfo {volume}
  {89}},\ \bibinfo {pages} {075302} (\bibinfo {year} {2014})}.

\bibitem{jiang}%
  \bibfield  {author} {\bibinfo {author} {\bibfnamefont {L.}~\bibnamefont
  {Jiang}}, \bibinfo {author} {\bibfnamefont {C.~H.}\ \bibnamefont {Yang}},
  \bibinfo {author} {\bibfnamefont {Z.}~\bibnamefont {Pan}}, \bibinfo {author}
  {\bibfnamefont {A.}~\bibnamefont {Rossi}}, \bibinfo {author} {\bibfnamefont
  {A.~S.}\ \bibnamefont {Dzurak}}, \ and\ \bibinfo {author} {\bibfnamefont
  {D.}~\bibnamefont {Culcer}},\ }\href {\doibase 10.1103/PhysRevB.88.085311}
  {\bibfield  {journal} {\bibinfo  {journal} {Phys. Rev. B}\ }\textbf {\bibinfo
  {volume} {88}},\ \bibinfo {pages} {085311} (\bibinfo {year}
  {2013})}.

\bibitem{genetic_design}%
  \bibfield  {author} {\bibinfo {author} {\bibfnamefont {L.}~\bibnamefont
  {Zhang}}, \bibinfo {author} {\bibfnamefont {J.-W.}\ \bibnamefont {Luo}},
  \bibinfo {author} {\bibfnamefont {A.}~\bibnamefont {Saraiva}}, \bibinfo
  {author} {\bibfnamefont {B.}~\bibnamefont {Koiller}}, \ and\ \bibinfo
  {author} {\bibfnamefont {A.}~\bibnamefont {Zunger}},\ }\href {\doibase
  10.1038/ncomms3396} {\bibfield  {journal} {\bibinfo  {journal} {Nat.
  Commun.}\ }\textbf {\bibinfo {volume} {4}},\ \bibinfo {pages} {2396}
  (\bibinfo {year} {2013})}.

\bibitem{dusko}%
  \bibfield  {author} {\bibinfo {author} {\bibfnamefont {A.}~\bibnamefont
  {Dusko}}, \bibinfo {author} {\bibfnamefont {A.~L.}\ \bibnamefont {Saraiva}},
  \ and\ \bibinfo {author} {\bibfnamefont {B.}~\bibnamefont {Koiller}},\
  }\href {\doibase 10.1103/PhysRevB.89.205307} {\bibfield  {journal} {\bibinfo
  {journal} {Phys. Rev. B}\ }\textbf {\bibinfo {volume} {89}},\ \bibinfo
  {pages} {205307} (\bibinfo {year} {2014})}.

\bibitem{xiao:032103}%
  \bibfield  {author} {\bibinfo {author} {\bibfnamefont {M.}~\bibnamefont
  {Xiao}}, \bibinfo {author} {\bibfnamefont {M.~G.}\ \bibnamefont {House}}, \
  and\ \bibinfo {author} {\bibfnamefont {H.~W.}\ \bibnamefont {Jiang}},\ }\href
  {\doibase 10.1063/1.3464324} {\bibfield  {journal} {\bibinfo  {journal}
  {Appl. Phys. Lett.}\ }\textbf {\bibinfo {volume} {97}},\ \bibinfo {eid}
  {032103} (\bibinfo {year} {2010})}.

\bibitem{borselli:123118}%
  \bibfield  {author} {\bibinfo {author} {\bibfnamefont {M.~G.}\ \bibnamefont
  {Borselli}}, \bibinfo {author} {\bibfnamefont {R.~S.}\ \bibnamefont {Ross}},
  \bibinfo {author} {\bibfnamefont {A.~A.}\ \bibnamefont {Kiselev}}, \bibinfo
  {author} {\bibfnamefont {E.~T.}\ \bibnamefont {Croke}}, \bibinfo {author}
  {\bibfnamefont {K.~S.}\ \bibnamefont {Holabird}}, \bibinfo {author}
  {\bibfnamefont {P.~W.}\ \bibnamefont {Deelman}}, \bibinfo {author}
  {\bibfnamefont {L.~D.}\ \bibnamefont {Warren}}, \bibinfo {author}
  {\bibfnamefont {I.}~\bibnamefont {Alvarado-Rodriguez}}, \bibinfo {author}
  {\bibfnamefont {I.}~\bibnamefont {Milosavljevic}}, \bibinfo {author}
  {\bibfnamefont {F.~C.}\ \bibnamefont {Ku}}, \bibinfo {author} {\bibfnamefont
  {W.~S.}\ \bibnamefont {Wong}}, \bibinfo {author} {\bibfnamefont {A.~E.}\
  \bibnamefont {Schmitz}}, \bibinfo {author} {\bibfnamefont {M.}~\bibnamefont
  {Sokolich}}, \bibinfo {author} {\bibfnamefont {M.~F.}\ \bibnamefont {Gyure}},
  \ and\ \bibinfo {author} {\bibfnamefont {A.~T.}\ \bibnamefont {Hunter}},\
  }\href {\doibase 10.1063/1.3569717} {\bibfield  {journal} {\bibinfo
  {journal} {Appl. Phys. Lett.}\ }\textbf {\bibinfo {volume} {98}},\
  \bibinfo {eid} {123118} (\bibinfo {year} {2011}{\natexlab{a}})}.

\bibitem{borselli:063109}%
  \bibfield  {author} {\bibinfo {author} {\bibfnamefont {M.~G.}\ \bibnamefont
  {Borselli}}, \bibinfo {author} {\bibfnamefont {K.}~\bibnamefont {Eng}},
  \bibinfo {author} {\bibfnamefont {E.~T.}\ \bibnamefont {Croke}}, \bibinfo
  {author} {\bibfnamefont {B.~M.}\ \bibnamefont {Maune}}, \bibinfo {author}
  {\bibfnamefont {B.}~\bibnamefont {Huang}}, \bibinfo {author} {\bibfnamefont
  {R.~S.}\ \bibnamefont {Ross}}, \bibinfo {author} {\bibfnamefont {A.~A.}\
  \bibnamefont {Kiselev}}, \bibinfo {author} {\bibfnamefont {P.~W.}\
  \bibnamefont {Deelman}}, \bibinfo {author} {\bibfnamefont {I.}~\bibnamefont
  {Alvarado-Rodriguez}}, \bibinfo {author} {\bibfnamefont {A.~E.}\ \bibnamefont
  {Schmitz}}, \bibinfo {author} {\bibfnamefont {M.}~\bibnamefont {Sokolich}},
  \bibinfo {author} {\bibfnamefont {K.~S.}\ \bibnamefont {Holabird}}, \bibinfo
  {author} {\bibfnamefont {T.~M.}\ \bibnamefont {Hazard}}, \bibinfo {author}
  {\bibfnamefont {M.~F.}\ \bibnamefont {Gyure}}, \ and\ \bibinfo {author}
  {\bibfnamefont {A.~T.}\ \bibnamefont {Hunter}},\ }\href {\doibase
  10.1063/1.3623479} {\bibfield  {journal} {\bibinfo  {journal} {Appl.
  Phys. Lett.}\ }\textbf {\bibinfo {volume} {99}},\ \bibinfo {eid} {063109}
  (\bibinfo {year} {2011}{\natexlab{b}})}.

\bibitem{lim}%
  \bibfield  {author} {\bibinfo {author} {\bibfnamefont {W.~H.}\ \bibnamefont
  {Lim}}, \bibinfo {author} {\bibfnamefont {C.~H.}\ \bibnamefont {Yang}},
  \bibinfo {author} {\bibfnamefont {F.~A.}\ \bibnamefont {Zwanenburg}}, \ and\
  \bibinfo {author} {\bibfnamefont {A.~S.}\ \bibnamefont {Dzurak}},\ }\href
  {http://stacks.iop.org/0957-4484/22/i=33/a=335704} {\bibfield  {journal}
  {\bibinfo  {journal} {Nanotechnology}\ }\textbf {\bibinfo {volume} {22}},\
  \bibinfo {pages} {335704} (\bibinfo {year} {2011})}.

\bibitem{simmons}%
  \bibfield  {author} {\bibinfo {author} {\bibfnamefont {C.~B.}\ \bibnamefont
  {Simmons}}, \bibinfo {author} {\bibfnamefont {J.~R.}\ \bibnamefont {Prance}},
  \bibinfo {author} {\bibfnamefont {B.~J.}\ \bibnamefont {Van~Bael}}, \bibinfo
  {author} {\bibfnamefont {T.~S.}\ \bibnamefont {Koh}}, \bibinfo {author}
  {\bibfnamefont {Z.}~\bibnamefont {Shi}}, \bibinfo {author} {\bibfnamefont
  {D.~E.}\ \bibnamefont {Savage}}, \bibinfo {author} {\bibfnamefont {M.~G.}\
  \bibnamefont {Lagally}}, \bibinfo {author} {\bibfnamefont {R.}~\bibnamefont
  {Joynt}}, \bibinfo {author} {\bibfnamefont {M.}~\bibnamefont {Friesen}},
  \bibinfo {author} {\bibfnamefont {S.~N.}\ \bibnamefont {Coppersmith}}, \ and\
  \bibinfo {author} {\bibfnamefont {M.~A.}\ \bibnamefont {Eriksson}},\ }\href
  {\doibase 10.1103/PhysRevLett.106.156804} {\bibfield  {journal} {\bibinfo
  {journal} {Phys. Rev. Lett.}\ }\textbf {\bibinfo {volume} {106}},\ \bibinfo
  {pages} {156804} (\bibinfo {year} {2011})}.

\bibitem{PhysRevB.86.115319}%
  \bibfield  {author} {\bibinfo {author} {\bibfnamefont {C.~H.}\ \bibnamefont
  {Yang}}, \bibinfo {author} {\bibfnamefont {W.~H.}\ \bibnamefont {Lim}},
  \bibinfo {author} {\bibfnamefont {N.~S.}\ \bibnamefont {Lai}}, \bibinfo
  {author} {\bibfnamefont {A.}~\bibnamefont {Rossi}}, \bibinfo {author}
  {\bibfnamefont {A.}~\bibnamefont {Morello}}, \ and\ \bibinfo {author}
  {\bibfnamefont {A.~S.}\ \bibnamefont {Dzurak}},\ }\href {\doibase
  10.1103/PhysRevB.86.115319} {\bibfield  {journal} {\bibinfo  {journal} {Phys.
  Rev. B}\ }\textbf {\bibinfo {volume} {86}},\ \bibinfo {pages} {115319}
  (\bibinfo {year} {2012})}.

\bibitem{yang2013}%
  \bibfield  {author} {\bibinfo {author} {\bibfnamefont {C.~H.}\ \bibnamefont
  {Yang}}, \bibinfo {author} {\bibfnamefont {A.}~\bibnamefont {Rossi}},
  \bibinfo {author} {\bibfnamefont {R.}~\bibnamefont {Ruskov}}, \bibinfo
  {author} {\bibfnamefont {N.~S.}\ \bibnamefont {Lai}}, \bibinfo {author}
  {\bibfnamefont {F.~A.}\ \bibnamefont {Mohiyaddin}}, \bibinfo {author}
  {\bibfnamefont {S.}~\bibnamefont {Lee}}, \bibinfo {author} {\bibfnamefont
  {C.}~\bibnamefont {Tahan}}, \bibinfo {author} {\bibfnamefont
  {G.}~\bibnamefont {Klimeck}}, \bibinfo {author} {\bibfnamefont
  {A.}~\bibnamefont {Morello}}, \ and\ \bibinfo {author} {\bibfnamefont
  {A.~S.}\ \bibnamefont {Dzurak}},\ }\href {\doibase
  10.1038/ncomms3069} {\bibfield  {journal} {\bibinfo
  {journal} {Nat. Commun.}\ }\textbf {\bibinfo {volume} {4}},\ \bibinfo
  {pages} {2069} (\bibinfo {year} {2013})}.

\bibitem{PhysRevLett.111.046801}%
  \bibfield  {author} {\bibinfo {author} {\bibfnamefont {K.}~\bibnamefont
  {Wang}}, \bibinfo {author} {\bibfnamefont {C.}~\bibnamefont {Payette}},
  \bibinfo {author} {\bibfnamefont {Y.}~\bibnamefont {Dovzhenko}}, \bibinfo
  {author} {\bibfnamefont {P.~W.}\ \bibnamefont {Deelman}}, \ and\ \bibinfo
  {author} {\bibfnamefont {J.~R.}\ \bibnamefont {Petta}},\ }\href {\doibase
  10.1103/PhysRevLett.111.046801} {\bibfield  {journal} {\bibinfo  {journal}
  {Phys. Rev. Lett.}\ }\textbf {\bibinfo {volume} {111}},\ \bibinfo {pages}
  {046801} (\bibinfo {year} {2013})}.

\bibitem{kott}%
  \bibfield  {author} {\bibinfo {author} {\bibfnamefont {T.~M.}\ \bibnamefont
  {Kott}}, \bibinfo {author} {\bibfnamefont {B.}~\bibnamefont {Hu}}, \bibinfo
  {author} {\bibfnamefont {S.~H.}\ \bibnamefont {Brown}}, \ and\ \bibinfo
  {author} {\bibfnamefont {B.~E.}\ \bibnamefont {Kane}},\ }\href {\doibase
  10.1103/PhysRevB.89.041107} {\bibfield  {journal} {\bibinfo  {journal}
  {Phys. Rev. B}\ }\textbf {\bibinfo {volume} {89}},\ \bibinfo {pages}
  {041107} (\bibinfo {year} {2014})}.

\bibitem{roche}%
  \bibfield  {author} {\bibinfo {author} {\bibfnamefont {B.}~\bibnamefont
  {Roche}}, \bibinfo {author} {\bibfnamefont {E.}~\bibnamefont
  {Dupont-Ferrier}}, \bibinfo {author} {\bibfnamefont {B.}~\bibnamefont
  {Voisin}}, \bibinfo {author} {\bibfnamefont {M.}~\bibnamefont {Cobian}},
  \bibinfo {author} {\bibfnamefont {X.}~\bibnamefont {Jehl}}, \bibinfo {author}
  {\bibfnamefont {R.}~\bibnamefont {Wacquez}}, \bibinfo {author} {\bibfnamefont
  {M.}~\bibnamefont {Vinet}}, \bibinfo {author} {\bibfnamefont {Y.-M.}\
  \bibnamefont {Niquet}}, \ and\ \bibinfo {author} {\bibfnamefont
  {M.}~\bibnamefont {Sanquer}},\ }\href {\doibase
  10.1103/PhysRevLett.108.206812} {\bibfield  {journal} {\bibinfo  {journal}
  {Phys. Rev. Lett.}\ }\textbf {\bibinfo {volume} {108}},\ \bibinfo {pages}
  {206812} (\bibinfo {year} {2012})}.

\bibitem{yuan}%
  \bibfield  {author} {\bibinfo {author} {\bibfnamefont {M.}~\bibnamefont
  {Yuan}},
  \bibinfo {author} {\bibfnamefont {R.}~\bibnamefont {Joynt}},
  \bibinfo {author} {\bibfnamefont {Z.}~\bibnamefont {Yang}},
  \bibinfo {author} {\bibfnamefont {C.}~\bibnamefont {Tang}},
  \bibinfo {author} {\bibfnamefont {D.~E.}\ \bibnamefont {Savage}},
  \bibinfo {author} {\bibfnamefont {M.~G.}\ \bibnamefont {Lagally}},
  \bibinfo {author} {\bibfnamefont {M.~A.}\ \bibnamefont {Eriksson}},
  \ and\
  \bibinfo {author} {\bibfnamefont {A.~J.}\ \bibnamefont {Rimberg}},\ }
  \href {\doibase 10.1103/PhysRevB.90.035302}
  {\bibfield{journal} {\bibinfo {journal} {Phys. Rev. B}\ }
  \textbf {\bibinfo {volume} {90}},\ \bibinfo {pages} {035302}
  (\bibinfo {year} {2014})}.

\bibitem{yamahata}%
  \bibfield  {author} {\bibinfo {author} {\bibfnamefont {G.}~\bibnamefont
  {Yamahata}}, \bibinfo {author} {\bibfnamefont {T.}~\bibnamefont {Kodera}},
  \bibinfo {author} {\bibfnamefont {H.~O.~H.}\ \bibnamefont {Churchill}},
  \bibinfo {author} {\bibfnamefont {K.}~\bibnamefont {Uchida}}, \bibinfo
  {author} {\bibfnamefont {C.~M.}\ \bibnamefont {Marcus}}, \ and\ \bibinfo
  {author} {\bibfnamefont {S.}~\bibnamefont {Oda}},\ }\href {\doibase
  10.1103/PhysRevB.86.115322} {\bibfield  {journal} {\bibinfo  {journal} {Phys.
  Rev. B}\ }\textbf {\bibinfo {volume} {86}},\ \bibinfo {pages} {115322}
  (\bibinfo {year} {2012})}.

\bibitem{hao}%
  \bibfield  {author} {\bibinfo {author} {\bibfnamefont {X.}~\bibnamefont
  {Hao}}, \bibinfo {author} {\bibfnamefont {R.}~\bibnamefont {Ruskov}},
  \bibinfo {author} {\bibfnamefont {M.}~\bibnamefont {Xiao}}, \bibinfo {author}
  {\bibfnamefont {C.}~\bibnamefont {Tahan}}, \ and\ \bibinfo {author}
  {\bibfnamefont {H.}~\bibnamefont {Jiang}},\ }\href {\doibase 10.1038/ncomms4860} {\bibfield
  {journal} {\bibinfo  {journal} {Nat. Commun.}\ }\textbf {\bibinfo {volume}
  {5}},\ \bibinfo{pages} {3860}  (\bibinfo {year} {2014})}.

\bibitem{goswami}%
  \bibfield  {author} {\bibinfo {author} {\bibfnamefont {S.}~\bibnamefont
  {Goswami}}, \bibinfo {author} {\bibfnamefont {K.~A.}\ \bibnamefont
  {Slinker}}, \bibinfo {author} {\bibfnamefont {M.}~\bibnamefont {Friesen}},
  \bibinfo {author} {\bibfnamefont {L.~M.}\ \bibnamefont {McGuire}}, \bibinfo
  {author} {\bibfnamefont {J.~L.}\ \bibnamefont {Truitt}}, \bibinfo {author}
  {\bibfnamefont {C.}~\bibnamefont {Tahan}}, \bibinfo {author} {\bibfnamefont
  {L.~J.}\ \bibnamefont {Klein}}, \bibinfo {author} {\bibfnamefont {J.~O.}\
  \bibnamefont {Chu}}, \bibinfo {author} {\bibfnamefont {P.~M.}\ \bibnamefont
  {Mooney}}, \bibinfo {author} {\bibfnamefont {D.~W.}\ \bibnamefont {van~der
  Weide}}, \bibinfo {author} {\bibfnamefont {R.}~\bibnamefont {Joynt}},
  \bibinfo {author} {\bibfnamefont {S.~N.}\ \bibnamefont {Coppersmith}}, \ and\
  \bibinfo {author} {\bibfnamefont {M.~A.}\ \bibnamefont {Eriksson}},\ }\href
  {\doibase 10.1038/nphys475} {\bibfield  {journal} {\bibinfo  {journal}
  {Nature Phys.}\ }\textbf {\bibinfo {volume} {3}},\ \bibinfo {pages} {41}
  (\bibinfo {year} {2007})}.
  
\bibitem{eriksson}%
  \bibfield  {author} {\bibinfo {author} {\bibfnamefont {M.~A.}\ \bibnamefont
  {Eriksson}}, \bibinfo {author} {\bibfnamefont {M.}~\bibnamefont {Friesen}},
  \bibinfo {author} {\bibfnamefont {S.~N.}\ \bibnamefont {Coppersmith}},
  \bibinfo {author} {\bibfnamefont {R.}~\bibnamefont {Joynt}}, \bibinfo
  {author} {\bibfnamefont {L.~J.}\ \bibnamefont {Klein}}, \bibinfo {author}
  {\bibfnamefont {K.}~\bibnamefont {Slinker}}, \bibinfo {author} {\bibfnamefont
  {C.}~\bibnamefont {Tahan}}, \bibinfo {author} {\bibfnamefont {P.~M.}\
  \bibnamefont {Mooney}}, \bibinfo {author} {\bibfnamefont {J.~O.}\
  \bibnamefont {Chu}}, \ and\ \bibinfo {author} {\bibfnamefont {S.~J.}\
  \bibnamefont {Koester}},\ }\href {\doibase 10.1007/s11128-004-2224-z}
  {\bibfield  {journal} {\bibinfo  {journal} {Quantum Inf. Process.}\
  }\textbf {\bibinfo {volume} {3}},\ \bibinfo {pages} {133} (\bibinfo {year}
  {2004})}.

\bibitem{trauzettel}%
  \bibfield  {author} {\bibinfo {author} {\bibfnamefont {B.}~\bibnamefont
  {Trauzettel}}, \bibinfo {author} {\bibfnamefont {D.~V.}\ \bibnamefont
  {Bulaev}}, \bibinfo {author} {\bibfnamefont {D.}~\bibnamefont {Loss}}, \ and\
  \bibinfo {author} {\bibfnamefont {G.}~\bibnamefont {Burkard}},\ }\href
  {\doibase 10.1038/nphys544} {\bibfield  {journal} {\bibinfo  {journal}
  {Nat. Phys.}\ }\textbf {\bibinfo {volume} {3}},\ \bibinfo {pages} {192}
  (\bibinfo {year} {2007})}.

\bibitem{maune}%
  \bibfield  {author} {\bibinfo {author} {\bibfnamefont {B.~M.}\ \bibnamefont
  {Maune}}, \bibinfo {author} {\bibfnamefont {M.~G.}\ \bibnamefont {Borselli}},
  \bibinfo {author} {\bibfnamefont {B.}~\bibnamefont {Huang}}, \bibinfo
  {author} {\bibfnamefont {T.~D.}\ \bibnamefont {Ladd}}, \bibinfo {author}
  {\bibfnamefont {P.~W.}\ \bibnamefont {Deelman}}, \bibinfo {author}
  {\bibfnamefont {K.~S.}\ \bibnamefont {Holabird}}, \bibinfo {author}
  {\bibfnamefont {A.~A.}\ \bibnamefont {Kiselev}}, \bibinfo {author}
  {\bibfnamefont {I.}~\bibnamefont {Alvarado-Rodriguez}}, \bibinfo {author}
  {\bibfnamefont {R.~S.}\ \bibnamefont {Ross}}, \bibinfo {author}
  {\bibfnamefont {A.~E.}\ \bibnamefont {Schmitz}}, \bibinfo {author}
  {\bibfnamefont {M.}~\bibnamefont {Sokolich}}, \bibinfo {author}
  {\bibfnamefont {C.~A.}\ \bibnamefont {Watson}}, \bibinfo {author}
  {\bibfnamefont {M.~F.}\ \bibnamefont {Gyure}}, \ and\ \bibinfo {author}
  {\bibfnamefont {A.~T.}\ \bibnamefont {Hunter}},\ }\href {\doibase
  10.1038/nature10707} {\bibfield  {journal} {\bibinfo  {journal} {Nature (London)}\
  }\textbf {\bibinfo {volume} {481}},\ \bibinfo {pages} {344} (\bibinfo {year}
  {2012})}.

\bibitem{PhysRevLett.106.086801}%
  \bibfield  {author} {\bibinfo {author} {\bibfnamefont {A.}~\bibnamefont
  {P\'alyi}}\ and\ \bibinfo {author} {\bibfnamefont {G.}~\bibnamefont
  {Burkard}},\ }\href {\doibase 10.1103/PhysRevLett.106.086801} {\bibfield  {journal} {\bibinfo  {journal} {Phys.
  Rev. Lett.}\ }\textbf {\bibinfo {volume} {106}},\ \bibinfo {pages} {086801}
  (\bibinfo {year} {2011})}.

\bibitem{PhysRevB.84.195463}%
  \bibfield  {author} {\bibinfo {author} {\bibfnamefont {G.~Y.}\ \bibnamefont
  {Wu}}, \bibinfo {author} {\bibfnamefont {N.-Y.}\ \bibnamefont {Lue}}, \ and\
  \bibinfo {author} {\bibfnamefont {L.}~\bibnamefont {Chang}},\ }\href
  {\doibase 10.1103/PhysRevB.84.195463} {\bibfield  {journal} {\bibinfo
  {journal} {Phys. Rev. B}\ }\textbf {\bibinfo {volume} {84}},\ \bibinfo
  {pages} {195463} (\bibinfo {year} {2011})}.

\bibitem{PhysRevB.88.125422}%
  \bibfield  {author} {\bibinfo {author} {\bibfnamefont {G.~Y.}\ \bibnamefont
  {Wu}}, \bibinfo {author} {\bibfnamefont {N.-Y.}\ \bibnamefont {Lue}}, \ and\
  \bibinfo {author} {\bibfnamefont {Y.-C.}\ \bibnamefont {Chen}},\ }\href
  {\doibase 10.1103/PhysRevB.88.125422} {\bibfield  {journal} {\bibinfo
  {journal} {Phys. Rev. B}\ }\textbf {\bibinfo {volume} {88}},\ \bibinfo
  {pages} {125422} (\bibinfo {year} {2013})}.

\bibitem{PhysRevB.82.205315}%
  \bibfield  {author} {\bibinfo {author} {\bibfnamefont {D.}~\bibnamefont
  {Culcer}}, \bibinfo {author} {\bibfnamefont {X.}~\bibnamefont {Hu}}, \ and\
  \bibinfo {author} {\bibfnamefont {S.}~\bibnamefont {Das~Sarma}},\ }\href
  {\doibase 10.1103/PhysRevB.82.205315} {\bibfield  {journal} {\bibinfo
  {journal} {Phys. Rev. B}\ }\textbf {\bibinfo {volume} {82}},\ \bibinfo
  {pages} {205315} (\bibinfo {year} {2010}{\natexlab{a}})}.

\bibitem{PhysRevLett.108.126804}%
  \bibfield  {author} {\bibinfo {author} {\bibfnamefont {D.}~\bibnamefont
  {Culcer}}, \bibinfo {author} {\bibfnamefont {A.~L.}\ \bibnamefont {Saraiva}},
  \bibinfo {author} {\bibfnamefont {B.}~\bibnamefont {Koiller}}, \bibinfo
  {author} {\bibfnamefont {X.}~\bibnamefont {Hu}}, \ and\ \bibinfo {author}
  {\bibfnamefont {S.}~\bibnamefont {Das~Sarma}},\ }\href {\doibase
  10.1103/PhysRevLett.108.126804} {\bibfield  {journal} {\bibinfo  {journal}
  {Phys. Rev. Lett.}\ }\textbf {\bibinfo {volume} {108}},\ \bibinfo {pages}
  {126804} (\bibinfo {year} {2012})}.

\bibitem{PhysRevB.86.035321}%
  \bibfield  {author} {\bibinfo {author} {\bibfnamefont {Y.}~\bibnamefont
  {Wu}}\ and\ \bibinfo {author} {\bibfnamefont {D.}~\bibnamefont {Culcer}},\
  }\href {\doibase 10.1103/PhysRevB.86.035321} {\bibfield  {journal} {\bibinfo
  {journal} {Phys. Rev. B}\ }\textbf {\bibinfo {volume} {86}},\ \bibinfo
  {pages} {035321} (\bibinfo {year} {2012})}.

\bibitem{kugel_khomskii_1973}%
  \BibitemOpen
  \bibfield  {author} {\bibinfo {author} {\bibfnamefont {K.~I.}\ \bibnamefont
  {Kugel}}\ and\ \bibinfo {author} {\bibfnamefont {D.~I.}\ \bibnamefont
  {Khomskii}},\ }\href@noop {} {\bibfield  {journal} {\bibinfo  {journal} {Zh.
  Eksp. Teor. Fiz}\ }\textbf {\bibinfo {volume} {64}},\ \bibinfo {pages} {1429}
  (\bibinfo {year} {1973})}.

\bibitem{kane_nature}%
  \bibfield  {author} {\bibinfo {author} {\bibfnamefont {B.~E.}\ \bibnamefont
  {Kane}},\ }\href {http://dx.doi.org/10.1038/30156} {\bibfield  {journal}
  {\bibinfo  {journal} {Nature (London)}\ }\textbf {\bibinfo {volume} {393}},\ \bibinfo
  {pages} {133} (\bibinfo {year} {1998})}.

\bibitem{vandersypen}%
  \bibfield  {author} {\bibinfo {author} {\bibfnamefont {L.~M.~K.}\
  \bibnamefont {Vandersypen}}, \bibinfo {author} {\bibfnamefont
  {R.}~\bibnamefont {Hanson}}, \bibinfo {author} {\bibfnamefont {L.~H.~W.}\
  \bibnamefont {van Beveren}}, \bibinfo {author} {\bibfnamefont {J.~M.}\
  \bibnamefont {Elzerman}}, \bibinfo {author} {\bibfnamefont {J.~S.}\
  \bibnamefont {Greidanus}}, \bibinfo {author} {\bibfnamefont {S.~D.}\
  \bibnamefont {{F}ranceschi}}, \ and\ \bibinfo {author} {\bibfnamefont
  {L.~P.}\ \bibnamefont {Kouwenhoven}},\ }in\ \href@noop {} {\emph {\bibinfo
  {booktitle} {Quantum Computing and Quantum Bits in Mesoscopic Systems}}}\
  (\bibinfo  {publisher} {Kluwer Academic/Plenum, New York},\ \bibinfo {year}
  {2003})\ \Eprint {http://arxiv.org/abs/arXiv:quant-ph/0207059}
  {arXiv:quant-ph/0207059}.

\bibitem{PhysRevB.79.085407}%
  \bibfield  {author} {\bibinfo {author} {\bibfnamefont {P.}~\bibnamefont
  {Recher}}, \bibinfo {author} {\bibfnamefont {J.}~\bibnamefont {Nilsson}},
  \bibinfo {author} {\bibfnamefont {G.}~\bibnamefont {Burkard}}, \ and\
  \bibinfo {author} {\bibfnamefont {B.}~\bibnamefont {Trauzettel}},\ }\href
  {\doibase 10.1103/PhysRevB.79.085407} {\bibfield  {journal} {\bibinfo
  {journal} {Phys. Rev. B}\ }\textbf {\bibinfo {volume} {79}},\ \bibinfo
  {pages} {085407} (\bibinfo {year} {2009})}.

\bibitem{burkard_imamoglu}%
  \bibfield  {author} {\bibinfo {author} {\bibfnamefont {G.}~\bibnamefont
  {Burkard}}\ and\ \bibinfo {author} {\bibfnamefont {A.}~\bibnamefont
  {Imamoglu}},\ }\href {\doibase 10.1103/PhysRevB.74.041307} {\bibfield
  {journal} {\bibinfo  {journal} {Phys. Rev. B}\ }\textbf {\bibinfo {volume}
  {74}},\ \bibinfo {pages} {041307} (\bibinfo {year} {2006})}.

\bibitem{Supplemental}
See Supplemental Material below,
which includes Ref. \cite{wardrop_doherty},
for details on the Schrieffer-Wolff transformation,
the time evolution with the effective Hamiltonian
including an analytical result for the phase $\alpha$ in Eq. (\ref{eqn:Fidelity}),
numerical results for the fidelity at the $m$th local maximum with $m = 1, 2, 3, 4$,
considerations on the influence of quasistatic noise,
and explicit sequences for the quantum gates.


\bibitem{wardrop_doherty}%
  \bibfield {author} {\bibinfo {author} {\bibnamefont {M.~P.}~\bibnamefont
  {Wardrop}}\ and\ \bibinfo {author} {\bibnamefont {A.~C.}~\bibnamefont {Doherty}},\ }
  \href {\doibase 10.1103/PhysRevB.90.045418} {\bibfield {journal}
  {\bibinfo {journal} {Phys. Rev. B}\ }\textbf {\bibinfo{volume} {90}},\
  \bibinfo {pages} {045418} (\bibinfo {year} {2014})}.

\bibitem{Pedersen200747}%
  \bibfield  {author} {\bibinfo {author} {\bibfnamefont {L.~H.}\ \bibnamefont
  {Pedersen}}, \bibinfo {author} {\bibfnamefont {N.~M.}\ \bibnamefont
  {M{\o}ller}}, \ and\ \bibinfo {author} {\bibfnamefont {K.}~\bibnamefont
  {M{\o}lmer}},\ }\href {\doibase
  10.1016/j.physleta.2007.02.069} {\bibfield  {journal}
  {\bibinfo  {journal} {Phys. Lett. A}\ }\textbf {\bibinfo {volume}
  {367}},\ \bibinfo {pages} {47 } (\bibinfo {year} {2007})}.

\bibitem{PhysRevLett110.146804}%
  \bibfield{author} {\bibinfo {author} {\bibfnamefont {O.~E}~\bibnamefont
  {Dial}},\ \bibinfo {author} {\bibfnamefont {M.~D.}~\bibnamefont
  {Shulman}},\ \bibinfo {author} {\bibfnamefont {S.~P.}~\bibnamefont
  {Harvey}},\ \bibinfo {author} {\bibfnamefont {H.}~\bibnamefont
  {Bluhm}},\ \bibinfo {author} {\bibfnamefont {V.}~\bibnamefont
  {Umansky}}, and\ \bibinfo {author} {\bibfnamefont {A.}~\bibnamefont
  {Yacoby}},\ }\href {\doibase 10.1103/PhysRevLett.110.146804} {\bibfield
  {journal} {\bibinfo  {journal} {Phys. Rev. Lett.}\ }\textbf {\bibinfo {volume}
  {110}},\ \bibinfo {pages} {146804} (\bibinfo {year} {2013})}.

\bibitem{APL102_232108}%
  \bibfield  {author} {\bibinfo {author} {\bibfnamefont {D.}~\bibnamefont
  {Culcer}} \ and \ \bibinfo {author} {\bibfnamefont {N.~M.}\ \bibnamefont {Zimmerman}},\ }
  \href {\doibase 10.1063/1.4810911} {\bibfield  {journal} {\bibinfo
  {journal} {Appl. Phys. Lett.}\ }\textbf {\bibinfo {volume} {102}},\ \bibinfo
  {pages} {232108} (\bibinfo {year} {2013})}.

\bibitem{divincenzo1995}%
  \bibfield  {author} {\bibinfo {author} {\bibfnamefont {D.~P.}\ \bibnamefont
  {DiVincenzo}},\ }\href {\doibase 10.1103/PhysRevA.51.1015} {\bibfield
  {journal} {\bibinfo  {journal} {Phys. Rev. A}\ }\textbf {\bibinfo {volume}
  {51}},\ \bibinfo {pages} {1015} (\bibinfo {year} {1995})}.

\bibitem{RevModPhys.54.437}%
  \bibfield  {author} {\bibinfo {author} {\bibfnamefont {T.}~\bibnamefont
  {Ando}}, \bibinfo {author} {\bibfnamefont {A.~B.}\ \bibnamefont {Fowler}}, \
  and\ \bibinfo {author} {\bibfnamefont {F.}~\bibnamefont {Stern}},\ }\href
  {\doibase 10.1103/RevModPhys.54.437} {\bibfield  {journal} {\bibinfo
  {journal} {Rev. Mod. Phys.}\ }\textbf {\bibinfo {volume} {54}},\ \bibinfo
  {pages} {437} (\bibinfo {year} {1982})}.

\bibitem{PhysRevB.61.2961}%
  \bibfield  {author} {\bibinfo {author} {\bibfnamefont {B.~E.}\ \bibnamefont
  {Kane}}, \bibinfo {author} {\bibfnamefont {N.~S.}\ \bibnamefont {McAlpine}},
  \bibinfo {author} {\bibfnamefont {A.~S.}\ \bibnamefont {Dzurak}}, \bibinfo
  {author} {\bibfnamefont {R.~G.}\ \bibnamefont {Clark}}, \bibinfo {author}
  {\bibfnamefont {G.~J.}\ \bibnamefont {Milburn}}, \bibinfo {author}
  {\bibfnamefont {H.~B.}\ \bibnamefont {Sun}}, \ and\ \bibinfo {author}
  {\bibfnamefont {H.}~\bibnamefont {Wiseman}},\ }\href {\doibase
  10.1103/PhysRevB.61.2961} {\bibfield  {journal} {\bibinfo  {journal} {Phys.
  Rev. B}\ }\textbf {\bibinfo {volume} {61}},\ \bibinfo {pages} {2961}
  (\bibinfo {year} {2000})}.

\bibitem{herring}%
  \bibfield  {author} {\bibinfo {author} {\bibfnamefont {C.}~\bibnamefont
  {Herring}},\ }\href {\doibase 10.1002/j.1538-7305.1955.tb01472.x} {\bibfield
  {journal} {\bibinfo  {journal} {Bell Syst. Tech. J.}\ }\textbf
  {\bibinfo {volume} {34}},\ \bibinfo {pages} {237} (\bibinfo {year}
  {1955})}.

\bibitem{PhysRev.101.944}%
  \bibfield  {author} {\bibinfo {author} {\bibfnamefont {C.}~\bibnamefont
  {Herring}}\ and\ \bibinfo {author} {\bibfnamefont {E.}~\bibnamefont {Vogt}},\
  }\href {\doibase 10.1103/PhysRev.101.944} {\bibfield  {journal} {\bibinfo
  {journal} {Phys. Rev.}\ }\textbf {\bibinfo {volume} {101}},\ \bibinfo {pages}
  {944} (\bibinfo {year} {1956})}.

\bibitem{PhysRevB.34.5621}%
  \bibfield  {author} {\bibinfo {author} {\bibfnamefont {C.~G.}\ \bibnamefont
  {Van~de Walle}}\ and\ \bibinfo {author} {\bibfnamefont {R.~M.}\ \bibnamefont
  {Martin}},\ }\href {\doibase 10.1103/PhysRevB.34.5621} {\bibfield  {journal}
  {\bibinfo  {journal} {Phys. Rev. B}\ }\textbf {\bibinfo {volume} {34}},\
  \bibinfo {pages} {5621} (\bibinfo {year} {1986})}.

\bibitem{people_bean}%
  \bibfield  {author} {\bibinfo {author} {\bibfnamefont {R.}~\bibnamefont
  {People}}\ and\ \bibinfo {author} {\bibfnamefont {J.~C.}\ \bibnamefont
  {Bean}},\ }\href {\doibase 10.1063/1.96499} {\bibfield  {journal} {\bibinfo
  {journal} {Appl. Phys. Lett.}\ }\textbf {\bibinfo {volume} {48}},\ \bibinfo
  {pages} {538} (\bibinfo {year} {1986})}.

\bibitem{PhysRevB.48.14276}%
  \bibfield  {author} {\bibinfo {author} {\bibfnamefont {M.~M.}\ \bibnamefont
  {Rieger}}\ and\ \bibinfo {author} {\bibfnamefont {P.}~\bibnamefont {Vogl}},\
  }\href {\doibase 10.1103/PhysRevB.48.14276} {\bibfield  {journal} {\bibinfo
  {journal} {Phys. Rev. B}\ }\textbf {\bibinfo {volume} {48}},\ \bibinfo
  {pages} {14276} (\bibinfo {year} {1993})}.

\bibitem{schaeffler}%
  \bibfield  {author} {\bibinfo {author} {\bibfnamefont {F.}~\bibnamefont
  {Sch\"affler}},\ }\href {http://stacks.iop.org/0268-1242/12/i=12/a=001}
  {\bibfield  {journal} {\bibinfo  {journal} {Semicond. Sci.
  Technol.}\ }\textbf {\bibinfo {volume} {12}},\ \bibinfo {pages} {1515}
  (\bibinfo {year} {1997})}.
  
\bibitem{APL95_0723102}%
  \bibfield  {author} {\bibinfo {author}
  {\bibfnamefont {D.}~\bibnamefont {Culcer}},
  \bibinfo {author} {\bibnamefont {X.}~\bibnamefont{Hu}},
  \ and \
  \bibinfo {author} {\bibnamefont {S.}~\bibnamefont{Das~Sarma}},\ }
  \href {http://dx.doi.org/10.1063/1.3194778}
  {\bibfield  {journal} {\bibinfo  {journal} {Appl. Phys. Lett.}\ }
  \textbf {\bibinfo {volume} {95}},\ \bibinfo {pages} {073102}
  (\bibinfo {year} {2009})}.

\bibitem{SciRep1_110}%
  \bibfield  {author} {\bibinfo {author}
  {\bibfnamefont {N.~S.}~\bibnamefont {Lai}},
  \bibinfo {author} {\bibnamefont {W.~H.}~\bibnamefont{Lim}},
  \bibinfo {author} {\bibnamefont {C.~H.}~\bibnamefont{Yang}},
  \bibinfo {author} {\bibnamefont {F.~A.}~\bibnamefont{Zwanenburg}},
  \bibinfo {author} {\bibnamefont {W.~A.}~\bibnamefont{Coish}},
  \bibinfo {author} {\bibnamefont {F.}~\bibnamefont{Qassemi}},
  \bibinfo {author} {\bibnamefont {A.}~\bibnamefont{Morello}},
  \ and \
  \bibinfo {author} {\bibnamefont {A.~S.}~\bibnamefont{Dzurak}},\ }
  \href {\doibase 10.1038/srep00110}
  {\bibfield  {journal} {\bibinfo  {journal} {Sci. Rep.}\ }
  \textbf {\bibinfo {volume} {1}},\ \bibinfo {pages} {110}
  (\bibinfo {year} {2011})}.

\end{thebibliography}
\end{document}